\documentclass[a4paper,11pt]{article}
\usepackage{jcappub} 
\usepackage{lineno}

\usepackage{multirow}

\usepackage{xcolor}
\usepackage[]{units}
\usepackage{amsmath} 

\usepackage[markup=underlined, color=red]{changes}
\setaddedmarkup{\textcolor{red}{#1}}

\arxivnumber{2501.12614} 
\title{\boldmath{Electric field reconstruction with three polarizations for the radio detection of ultra-high energy particles}}

\author[a,b]{Kewen Zhang,}
\author[c,d]{Tim Huege,}
\author[e,f,l]{Ramesh Koirala,}
\author[a,b]{Pengxiong Ma,}
\author[g,h]{Matías Tueros,}
\author[i]{Xin Xu,}
\author[e,f,j,k, 1]{Chao Zhang,}
\author[i]{Pengfei Zhang,}
\author[a,b,1] {Yi Zhang,\note{Corresponding author.}}
\affiliation[a]{
    {Key Laboratory of Dark Matter and Space Astronomy, Purple Mountain Observatory, Chinese Academy of Sciences}, No. 10 Yuanhua Road, Nanjing, 
    China}
\affiliation[b]{School of Astronomy and Space Science, University of Science and Technology of China, Hefei 230026, China}
\affiliation[c] {Institute for Astroparticle Physics (IAP), Karlsruhe Institute of Technology, Karlsruhe, Germany}
\affiliation[d] {Astrophysical Institute, Vrije Universiteit Brussels, Belgium}
\affiliation[e]{School of Astronomy and Space Science, Nanjing University, Nanjing 210023, China}
\affiliation[f]{Key laboratory of Modern Astronomy and Astrophysics, Nanjing University, Ministry of Education, Nanjing 210023, China}
\affiliation[g]{IFLP - CCT La Plata - CONICET, Casilla de Correo, La Plata, Argentina}
\affiliation[h]{Depto. de Física, Fac. de Cs. Ex., Universidad Nacional de La Plata, Casilla de Coreo, La Plata, Argentina}
\affiliation[i]{School of Electronic Engineering, Xidian University, No.2 South Taibai Road, Xi’an, China }
\affiliation[j] {Department of Astronomy, School of Physics, Peking University, Beijing 100871, China}
\affiliation[k]{Kavli institute for astronomy and astrophysics, Peking University, Beĳing, China}
\affiliation[l]{Space Research Centre, Faculty of Technology, Nepal Academy of Science and Technology (NAST), Khumaltar, Lalitpur, Nepal}
\emailAdd{chao.zhang@nju.edu.cn, zhangpf@mail.xidian.edu.cn, zhangyi@pmo.ac.cn}

\abstract{
Accurate reconstruction of the electric field produced by Extensive Air Showers from the signals recorded by the antennas is essential for the radio detection technique, as the key parameters needed to retrieve information about the primary particle that generated the shower are the amplitude, polarization, frequency spectrum and energy fluence carried by the electric field at each measurement position. Conventional electric field reconstruction methods primarily focus on antennas with two horizontal polarizations. In this paper, we introduce an analytical $\chi^2$ minimization method that operates with both two and three polarizations, providing the reconstructed electric field at each antenna. This solution has been verified for simple and realistic antenna responses, with a particular focus on inclined air showers. Our method achieves a standard deviation better than 4\% in determining the peak envelope amplitude of the electric field and better than 6\% in the estimation of the energy fluence, with an antenna response dependent bias. Additionally, we have studied the dependence of the method with arrival direction showing that it has a good performance in almost all of them. This work also demonstrates that incorporating vertically polarized antennas enhances the precision of reconstruction, leading to a more accurate and reliable electric field estimation for inclined air showers. Consequently, the method  improves our ability to extract information about cosmic rays from the detected signals in current and future experiments.
}


\begin{document}
\maketitle
\flushbottom

\section{Introduction}
\label{intro}
Cosmic rays with energies ranging from approximately 100 PeV to a few EeV mark a crucial transition in the energy spectrum, connecting Galactic and extragalactic origins. The “knee” at around \unit[4]{PeV} likely reflects the rigidity-dependent cutoff of Galactic accelerators, while the “ankle” near 1 EeV signifies the increasing dominance of extragalactic sources. When the energy exceeds 1 EeV, it is called ultra-high-energy cosmic rays (UHECRs). These particles possess energies beyond what laboratory accelerators can achieve, offering a unique opportunity to explore fundamental physics. Understanding the composition and spectrum of cosmic rays in this energy range  provides insights into the mechanisms behind their acceleration and the interplay between Galactic and extragalactic magnetic fields \cite{Bl_mer_2009, PierreAuger:2010gfm, Ackermann:2022rqc, Coleman_2023, zhang2024UHECR}.

The flux of high-energy cosmic rays generally follows a steeply falling power law \cite{Abreu_2021}, resulting in an exceedingly low flux at high energies. 
Detectors with a large effective area are needed to collect adequate statistics of UHECRs within a reasonable time frame. 

Ground-based particle detectors (PD), deployed over a large area, provide a well-established technique for the detection of UHECRs, by detecting the extensive air showers (EAS) generated by the interactions of UHECRs with Earth's atmosphere \cite{pierre2015pierre,abu2012surface,aartsen2017icecube}. Ground-based antenna arrays offer an alternative and complementary approach to detect these EAS through the measurement of the radio signals they produce \cite{Huege_2016,Schr_der_2017}.

Inclined air showers — those with zenith angles exceeding 65$^\circ$ — are of particular interest for UHE particle studies. On this type of events, particle detectors mostly measure the muon content of the particle cascade due to the atmospheric absorption of it's electromagnetic component. Radio detection (RD), in contrast, provides direct access to the electromagnetic component and can thus complement particle detectors, offering additional advantages such as low costs, a near-100\% duty cycle, and high precision in reconstructing energy, depth of shower maximum $X_\mathrm{max}$, and arrival direction. The combination of RD and PD is particularly beneficial for enhancing sensitivity to mass composition \cite{Apel_2014, Bezyazeekov_2018, Huege_2023, Schr_der_2023}.  

The radio emission from EAS is known to be induced by the geomagnetic and Askaryan effects. The geomagnetic effect arises from the separation of electrons and positrons in the Earth's geomagnetic field due to the Lorentz force in combination with their interactions in the atmosphere \cite{kahn1966radiation}. The separation generates a time-varying transverse current, leading to radio emission contributing the majority of the radiation energy.  The Askaryan effect \cite{1962JPSJS..17C.257A} also contributes to the radiated energy, normally at a level of less than 10\% depending on the atmospheric density. This effect occurs whenever high-energy charged particles traverse a dense medium, causing additional electrons from the medium to be swept into the shower front through Compton scattering or ionization while the positive ions stay behind, while the number of positrons decreases due to annihilation in the medium. This process creates a time-varying net charge excess at the shower front, resulting in radio emission. Recent analyses have shown a third emission mechanism under high magnetic fields and low atmospheric density, where the charged particles can be deflected over longer paths, causing the charged particles to gyrate and produce synchrotron-like emission, as well as general reduction of the emission due to the loss of coherence \cite{James:2022mea, PhysRevLett.132.231001}.

A key feature of the radio emission is a Cherenkov ring in the shower plane (the plane perpendicular to the direction of the primary particle)  generated due to non-unity refractive index of the atmosphere.\cite{PhysRevLett.107.061101}. 
This creates an elliptical region of maximum coherence in the observed signal on the ground, sometimes referred to as the Cherenkov ring. The energy fluence distribution on this ring is asymmetric, arising from the distinct orientations of the electric field vectors associated with the two primary emission mechanisms. The electric field induced by the transverse current effect aligns with the direction of the Lorentz force, while that from the Askaryan effect exhibits a radial symmetry pointing towards the shower axis. This results in different polarizations of the electric field with respect to the shower core. Additionally, for inclined air showers, the propagation length of the emission above the shower axis is much longer (signals arriving late at the antennas) than that below the shower axis (signals arriving early). As a result of this geometric early-late effect, extra asymmetries arise on the ground, which can be corrected for as shown in \cite{Huege_2019, Schluter:2020tdz}.

Simulation tools such as CoREAS \cite{Huege_2013} and ZHAireS \cite{Alvarez_Mu_iz_2012} are widely used for simulating radio emissions and study their properties. Experimental efforts, exemplified by pioneering projects like LOPES \cite{2005Natur_435_313F} and CODALEMA \cite{Ardouin_2005}, have demonstrated the feasibility of radio detection of EAS.  Following the successful validation of this detection technique, numerous  radio experiments have been launched. Some recent projects have further proposed the use of pure radio detection methods, optimized for various air shower geometries, including inclined and upward-going air showers, as well as in-ice particle cascades \cite{Huege_2016, Schr_der_2017, GRAND:2018iaj, abbasi2018depth, RNO-G:2020rmc}.

These radio-emission simulations and experiments paved the way for sophisticated reconstruction techniques. Sub-degree precision in direction reconstruction has been achieved by analyzing the radio signal's peak \cite{Apel_2014,Corstanje_2015,decoene2020sources}. For vertical air showers, an ultimate energy resolution of 3\% is expected to be achieved with the SKA \cite{Buitink:2023rso, corstanje2023} thanks to its high antenna density and increased bandwidth. For inclined showers, a resolution better than the 5$-$10\% level is expected from realistic end-to-end simulation studies on the basis of an analytical description of the radio-emission footprint, a parameterization of the signal and density corrections to the radiation energy \cite{Schlüter_2023}. Furthermore, variations in the radio footprint caused by different $X_\mathrm{max}$ values for a fixed incident direction and energy have been utilized for the reconstruction of $X_\mathrm{max}$, achieving resolutions better than 15 g/cm$^2$ at the highest energies in AERA \cite{PhysRevD.109.022002} and expecting resolutions better than 8 g/cm$^2$ in SKA with established techniques \cite{Buitink:2023rso, corstanje2023, Buitink_2014,Buitink_2016,Huege_2017,2024icrc.confE.503B,Corstanje:2025wbc}.

The reconstruction of these shower properties relies on accurate information about the electric field. However, antennas measure voltage signals, making it crucial to reconstruct the electric field as one the first steps for any subsequent analysis. The most widely-used technique for reconstructing the electric field from voltage traces, involving the inversion of a 2×2 response matrix,  was developed by AERA in \cite{Abreu_2011} and improved in \cite{Glaser_2019, welling2019reconstructing}, in which a forward-folding technique using multiple channel measurements was later developed to improve the reconstruction accuracy for signals with low SNR, for horizontal EAS, and for signals with bandwidths smaller than that of the detector. A more recent technique employs information field theory to reconstruct signals also with low SNR with high precision \cite{welling2021reconstructing,simon2024information}. These methods, however, have been applied primarily to antennas with two horizontal polarizations, leading to incomplete sampling of the three-dimensional electric field. For the case of three polarizations, the LOPES-3D experiment pioneered multiple weighted reconstruction methodologies, including polarization-direction-specific weighting factors calibrated according to their sensitivity to incident angles \cite{Huber2014_1000043289, huber2022lopes}, with no significant improvement relative to the conventional matrix inversion method.

In this context, we introduce a novel approach based on an analytical $\chi^2$ minimization for reconstructing the electric field. It incorporates not only the two horizontal polarizations but also the additional vertical polarization, allowing a more complete and precise reconstruction of the electric field without assumptions on the signal shape, giving a robust electric field estimation even under varying signal conditions and antenna responses.

This article is organized as follows: section 2 covers the simulations of radio emission, antenna response, and background noise used for reconstruction in our analysis. Section 3 introduces the analytical $\chi^2$ minimization method for electric field reconstruction using three polarizations. Section 4 evaluates the method’s performance in terms of reconstruction accuracy for the Hilbert envelope, energy fluence, and direction dependence, comparing two antenna types under realistic conditions. Section 5 presents the conclusions and outlook. Further technical details are given in several appendixes: Appendix A elaborates on the antenna response. Appendix B outlines the distribution of the relative error of the Hilbert peak envelope amplitude. Appendix C presents the Hilbert-transformed peak envelope amplitude computed using the matrix inversion method.

\section{Simulations}
\label{sims}
This work is based on detailed simulations of EAS and their radio emission. We first created a library of air showers with their associated electric fields. Next, we incorporated specific antenna models to convert the electric fields into voltage signals. Finally, we included a galactic noise model to account for background in the radio signals. By combining these elements, we generated simulated voltage data that closely matches experimental data, providing a robust foundation for evaluating the reconstruction of the electric field.

\subsection{Simulation of cosmic ray radio signals}
\label{sim_set}
We used the ZHAireS simulation package \cite{Alvarez_Mu_iz_2012} to simulate the radio emissions induced by EAS initiated by proton and iron nuclei. We created a Monte Carlo (MC) data set covering zenith angles ($\theta$)  from 63.0$^\circ$ to 87.1$^\circ$ in equidistant steps of $\log_{10}(1/\cos(\theta))=0.08$, focusing specifically on inclined EAS. The azimuth angle ($\phi$) spans from 0 to 180$^\circ$, with increments of 45$^\circ$. The primary particle energy $E_\mathrm{shower}$ ranges from \unit[0.126]{EeV} to \unit[3.98]{EeV} in logarithmic steps of 0.1. The simulation site is a radio-quiet area in Dunhuang, China. The geomagnetic field at this site has a strength of \unit[56]{$\mu $T} and an inclination angle of 61$^\circ$. For the atmospheric model in this study, we employ the extended Linsley US-standard atmosphere \cite{Heck:1998vt}. This model incorporates an exponential refractive index profile, characterized by a scale height of \unit[8.2]{km} and a sea-level refractive index of 1.000325. The simulated event data set is equally divided between proton and iron air showers. A total of 4,160 events were simulated to ensure statistical robustness in the analysis. 

\begin{figure}
    \centering
    \includegraphics[width=0.65\linewidth]{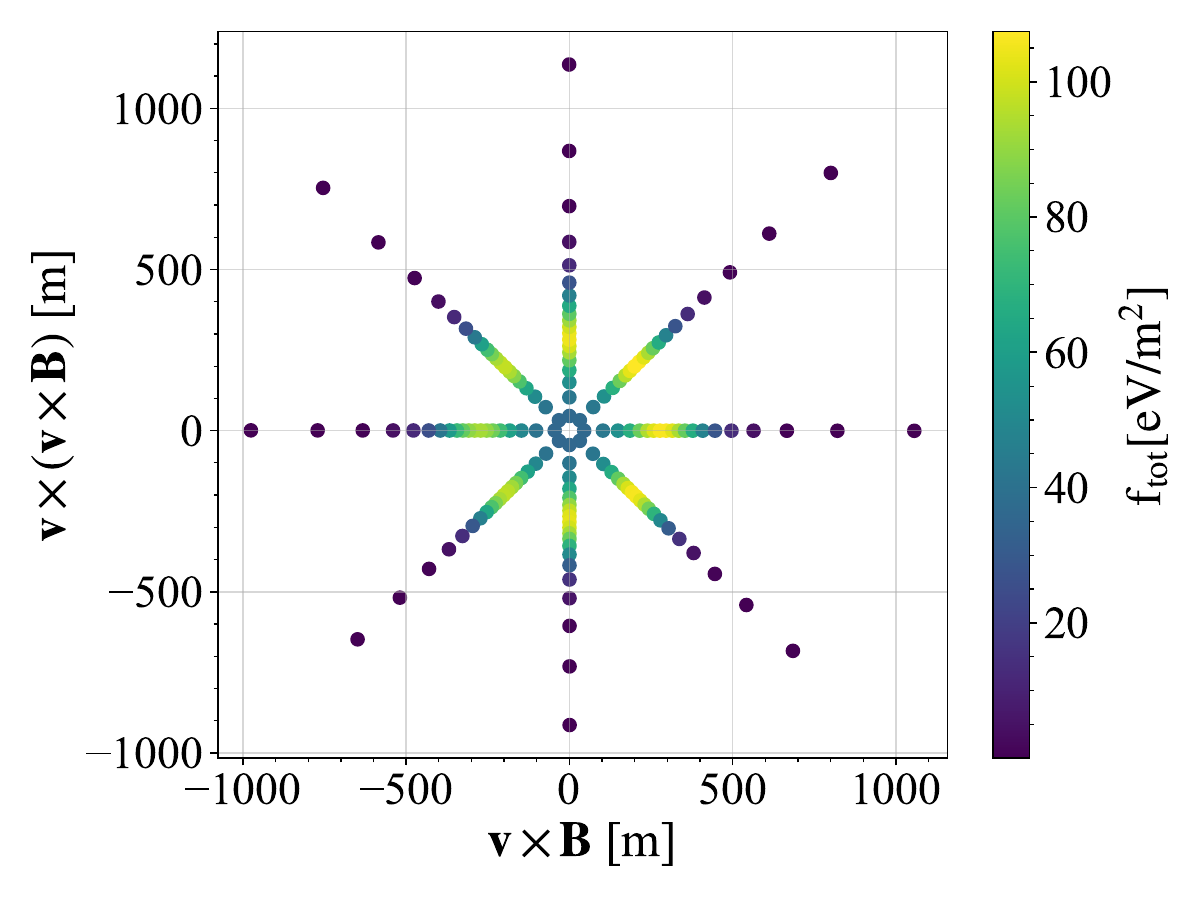}
    \caption{Example radio footprint of a simulated proton-induced air shower with a primary energy of \unit[0.631]{EeV}, zenith angle $\theta = 67.8^\circ$, and azimuth angle $\varphi = 45^\circ$, projected to the shower plane, with color indicating the strength of energy fluence.}
    \label{fig:footprint}
\end{figure}

For these simulations a star-shaped antenna array in the shower plane, as shown in Figure \ref{fig:footprint}, was projected to the ground plane to create an array consisting of 160 radio antennas, arranged in eight arms with 20 antennas per arm. The spacing between concentric rings was optimized to align with the Cherenkov angle, ensuring the peak signal was centrally positioned within the simulated lateral footprint. This geometric configuration was designed to guarantee complete coverage of the radio-emission footprint.

The components of the simulated electric field correspond to three polarization directions: $\mathrm{E_x}$ (South-North), $\mathrm{E_y}$ (East-West), and $\mathrm{E_z}$ (Down-Up). We selected a time binning of \unit[0.5]{ns} within a \unit[1000]{ns} time window,  yielding a frequency resolution of \unit[1]{MHz} after performing a Fourier transform.

\subsection{Antenna responses}
\label{ant_resp}
When using only horizontally aligned antennas, a fraction of horizontal events cannot be detected, as a significant portion of the signal can be vertically polarized. To enlarge the sky coverage of sparse arrays, it is imperative to include an additional vertical antenna arm \cite{Huber2014_1000043289}, or have two antennas that are partly sensitive to the vertical polarization component such as the dual-polarized SALLA antennas in the Auger Radio Detector and the LPDA antennas in the AERA radio array \cite{Huege_2023,Huege:2024nic}.

The study we present here tests two types of antennas for reconstructing the electric field with three polarizations: a simple three-arm dipole antenna and a more complex three-arm  HORIZON antenna \cite{GRAND:2018iaj, GRAND:2023mco}. The dipole antenna serves as a methodological baseline for developing the reconstruction method, whereas the HORIZON antenna is used for verification and performance evaluation. Antenna responses were simulated using Ansys HFSS \cite{ansysHFSS}, with all simulations incorporating ground effects to ensure accurate representation of real-world operational conditions. For a representative frequency, the antenna response patterns are shown in Appendix \ref{Responses}.

\subsubsection{Dipole antenna}
\label{3Ddipole}
The dipole antenna consists of three mutually perpendicular arms.  Each arms consist of two oscillators, each measuring \unit[1.3]{meters}, resulting in a total length of \unit[2.6]{meters} per arm. Antennas are positioned \unit[3]{meters} above the ground and operate within a frequency range of 30 to \unit[80]{MHz}, making it suitable for radio detection experiments.
The horizontal arms are aligned along the South-North (SN) and East-West (EW) directions, with the vertical arm representing the z polarization. In this context, the geomagnetic north is defined as the x-axis, and the east is the y-axis. Such a design was initially implemented in the LOPES-3D experiment \cite{huber2022lopes, Huber2014_1000043289}.

\begin{figure}[htbp!]
\centering
\includegraphics[width=1\textwidth]{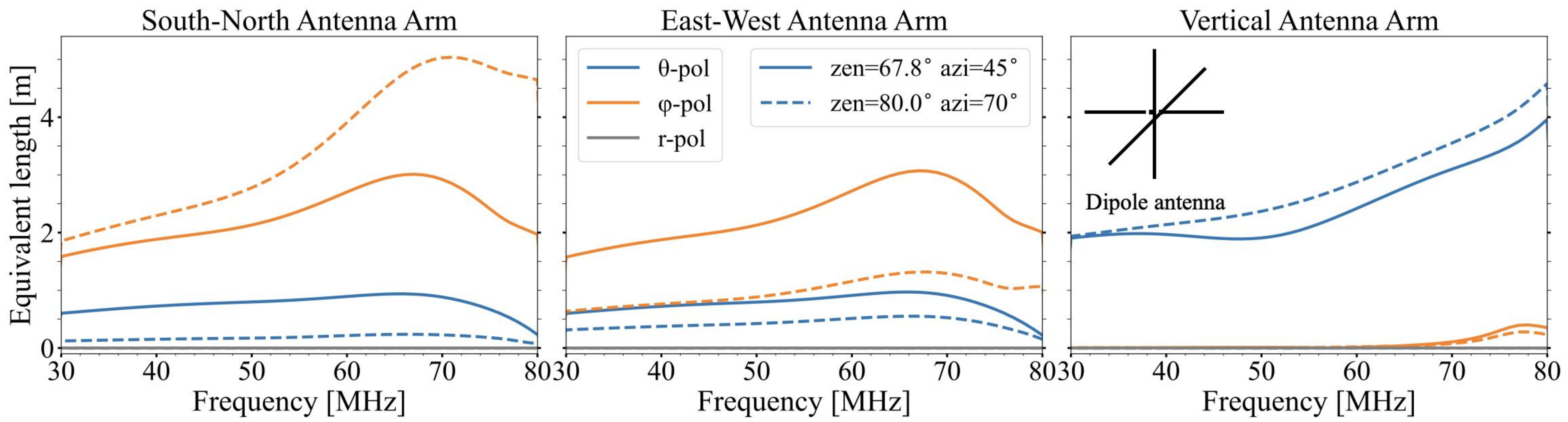}
    \caption{Frequency-domain response of a dipole antenna with three polarizations. The left, middle and right panels correspond to the North-South, East-West and vertical polarization arms, respectively. Solid and dashed lines denote different incident directions parametrized in spherical coordinates represented by $(\theta, \varphi, r)$.  The response in spherical representation, showing that the response in the radial direction (r) is close to zero, indicates the validity of the far-field approximation. A schematic of the antenna geometry  is shown in the upper left part of right panel.
    }
    \label{fig:ant_respn_dipole}
\end{figure}

Figure \ref{fig:ant_respn_dipole} illustrates the response of each antenna arm for two cases of zenith and azimuth angles. The polarizations are represented in spherical coordinates \((\theta, \varphi, r)\) \cite{abreu2012antennas}. In both cases, a single resonance is seen within the \unit[30-80]{MHz} range, leading to a relatively smooth variation across the frequency band, with only minor irregularities near the upper end.

\subsubsection{HORIZON antenna}
\label{sec:Horizon}

The HORIZON antenna consists of a symmetric butterfly-shape steel radiator with three mutually perpendicular antenna arms. The antenna arms are oriented as in a dipole antenna and operate within a frequency range of \unit[50-200]{MHz}. It consists of two radiators in the X- and Y-arms, and only one in the Z-arm. The antenna is equipped with an impedance-matching network. The HORIZON antenna design tries to optimize detection capability for inclined EAS while retaining sensitivity to less inclined showers. The design prioritizes phase-center consistency, though at the expense of omnidirectionality. This antenna type was inspired by the “butterfly antenna” used in the CODALEMA \cite{CHARRIER2012S142}, and it has been further refined and deployed in the GRAND experiment \cite{GRAND:2018iaj, GRAND:2023mco}. To ensure a fair comparison with the dipole antenna, the HORIZON antenna is also positioned 3 meters above the ground, which is 0.2 meters lower than the original design specified in \cite{GRAND:2024atu}. 

\begin{figure}[htbp]
\centering
\includegraphics[width=1\textwidth]{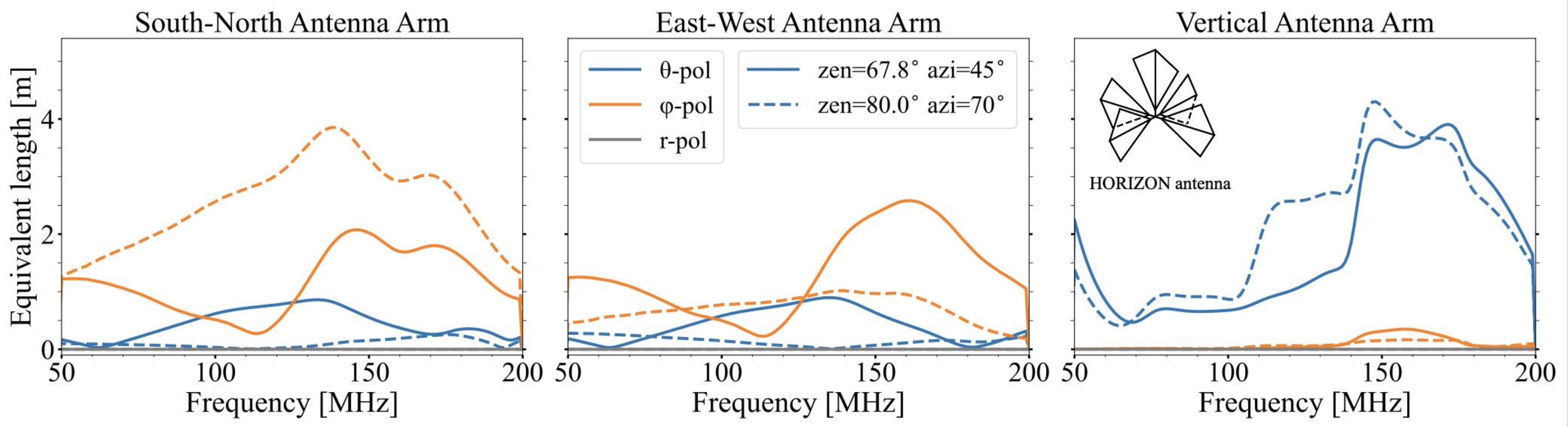}
\caption{Frequency-domain response of the HORIZON antenna is illustrated in the left to right panels, depicting the South-North, East-West and vertical antenna polarizations components, respectively. Solid and dashed lines denote different signal incident directions. A schematic of the antenna geometry is provided in the upper-left quadrant of the right-hand panel.}
\label{fig:ant_resp}
\end{figure}

Increasing  the bandwidth of the antenna introduces frequency-dependent lobes in the antenna beam. Despite the complexity in antenna response, this design achieves an efficient broadband interception of the incoming electric field energy from all directions. Figure \ref{fig:ant_resp} illustrates the response of the HORIZON antenna adopting the same coordinate conventions as in Figure \ref{fig:ant_respn_dipole}. Unlike the dipole antenna, this design exhibits multiple resonant frequencies. For instance, at a zenith angle of 67.8$^\circ$ and an azimuth angle of 45$^\circ$ in the East-West polarization, the $\varphi$-component displays two resonant frequencies, while the $\theta$-component shows negligible gain near \unit[60]{MHz} and \unit[180]{MHz}. More resonances are observed in the vertical polarization. This spectral complexity complicates electric field reconstruction — a challenge addressed in Section \ref{least_estm}.

\subsection{Electric field to voltage}
 Building on the antenna response presented in Section \ref{ant_resp}, we now relate the incident electric field to the open-circuit voltage via the vector equivalent length. The open-circuit voltage is the convolution of the vector equivalent length (VEL) of the antenna as shown in Figure \ref{fig:ant_respn_dipole} and \ref{fig:ant_resp}, with the simulated electric field, and is given by:
\begin{equation}
V(t)=\vec{H}(t) * \vec{E}(t)
\label{eq:v_t}
\end{equation}
where \(V(t)\) represents the measured voltage, \(\vec{H}(t)\) is the VEL, \(\vec{E}(t)\) is the electric field, and \(*\) denotes the convolution operation. For simplicity, analyses are typically conducted in the frequency domain after applying a Fourier transformation. In the frequency domain, the relationship corresponds to \cite{abreu2012antennas}:
\begin{equation}
    \mathcal{V}(f)=\vec{\mathcal{H}}(f)\cdot \vec{\mathcal{E}}(f)
    \label{eq:v_f}
\end{equation}
where $\mathcal{V}(f)$, $\vec{\mathcal{H}}(f)$ and $\vec{\mathcal{E}}(f)$ are  the voltage, VEL, and electric field, respectively, in the frequency domain following the Fourier transform (denoted in Italian font format).

\subsection{Background noise}
\label{bkg_noise}
The reconstruction of the electric field  presents significant challenges due to the background noise. To mitigate the impact of human activities such as radio telecommunications and navigation systems, radio detection experiment sites are chosen strategically to be located in remote areas. However, sporadic contamination from satellite and aircraft communications, as well as FM radio signals, can appear in the data. These sources typically occupy discrete frequencies or narrow spectral bands that  can be suppressed through targeted filtering and Radio Frequency Interference (RFI) mitigation protocols. Although transient phenomena such as thunderstorms, solar flares, or subterranean pre-seismic activity can generate complex background noise, their episodic nature excludes them from the scope of this investigation. In this scenario, the galactic radio background is considered to be the dominant irreducible noise source in the specified frequency band \cite{alvarez2020giant}. This type of noise, which marked the birth of radio astronomy, was first identified by Karl Jansky and dominantly originates from the diffuse synchrotron radiation within the Galaxy \cite{busken2023uncertainties}.

\begin{figure}[htbp]
\centering
\includegraphics[width=1\textwidth]{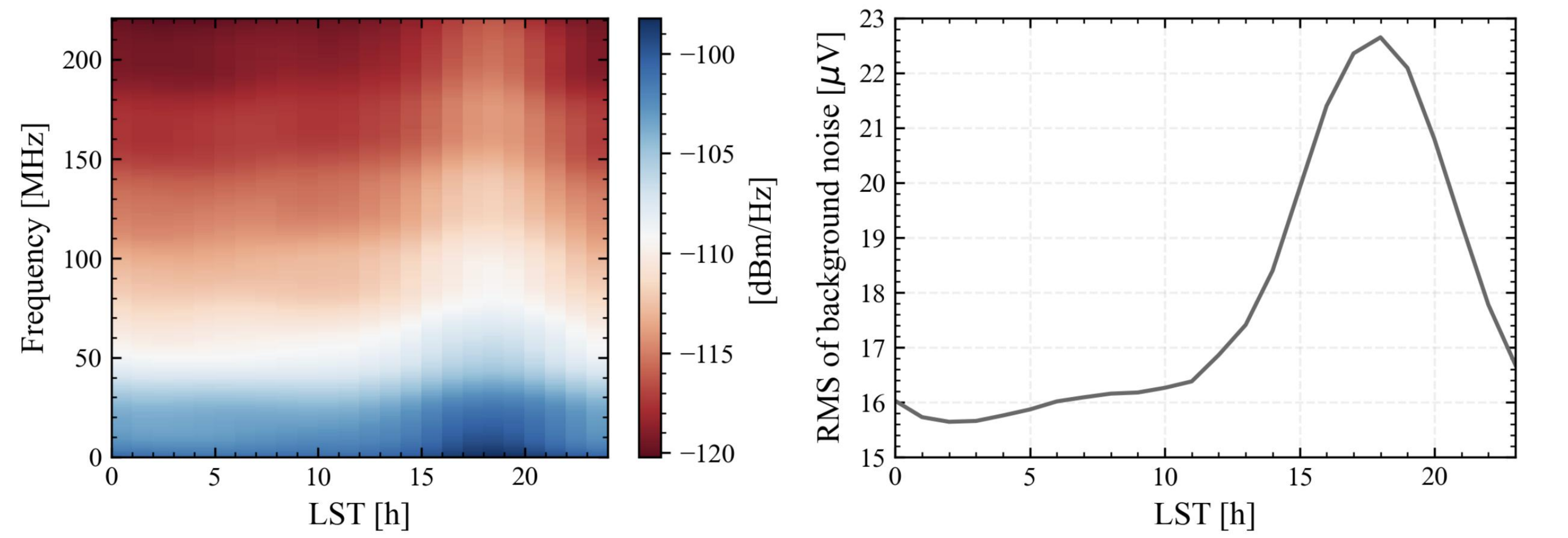}
    \caption{East-West polarization of the Galactic noise generated by the LFmap model. Left panel: power spectral density illustrating the variation of the Galactic noise with the LST, decreasing overall with frequency. Right panel: RMS of Galactic noise in the time domain as a function of the LST. }
      \label{fig:Gal_bkg}
\end{figure}

To model the background noise spectrum across the 30 to \unit[200]{MHz} band, we utilize the LFmap model \cite{polisensky2007lfmap}. This model incorporates multiple astrophysical components: the Cosmic Microwave Background (CMB), isotropic diffuse radiation, Galactic synchrotron emission, and extragalactic sources. The resulting sky temperature follows a power dependence: \begin{equation}
    T_\mathrm{sky}(f)\propto f^{-\beta}
    \label{eq:T_f}
\end{equation}
where $\beta$ denotes the spectral index, varying typically between 2.1 and 2.7 in the  \unit[45]{MHz} to \unit[408]{MHz} band \cite{Guzm_n_2010}, governed by the relative contributions of thermal bremsstrahlung and non-thermal synchrotron processes. The spectral brightness is derived from the sky temperature via the Rayleigh-Jeans approximation:  $B_\mathrm{f}=\frac{2k_\mathrm{B} f^2 T_\mathrm{sky}}{c^2}$, where $k_B$ is the Boltzmann constant, and $c$ is the speed of light. The sky noise power is then expressed as:

\begin{equation}
    P_\mathrm{sky}(t',f)=\frac{1}{2}\int B_f(\theta,\varphi,t') A_e(\theta,\varphi,f)\sin{\theta}\mathrm{d}\theta \mathrm{d}\varphi 
    \label{eq:P_intg}
\end{equation}
where $t'$ represents the the local sidereal time (LST), $A_e(\theta, \varphi, f)$ is effective aperture of the antenna (obtained from the antenna gain), and the antenna efficiency is assumed ideal ($\eta_A=1$). The effective aperture relates to antenna gain  through $A_e = \frac{\lambda^2}{4\pi}G$, where $G$ is the antenna gain, $G=1$ for a lossless resonant antenna under impedance-matched conditions. The left panel of Figure \ref{fig:Gal_bkg} illustrates this background power spectrum for the East-West polarization component.

In this analysis,  we omit specific antenna design details and assume a typical antenna impedance of $Z_0  = 50 \Omega$, corresponding to an ideal impedance-matched condition. In practice, antenna impedance varies with frequency and may deviate from this idealized assumption. However, this simplification provides a consistent noise level, thereby establishing a reproducible baseline for comparative analyses. With impedance matching, 50\% of the incident  power is transferred to the receiver. Consequently, the voltage induced by the Galactic background is therefore derived as:

\begin{equation}
    V_\mathrm{sky}=2\sqrt{Z_0 P_\mathrm{sky}}
    \label{eq:V_sky}
\end{equation}
The right panel of Figure \ref{fig:Gal_bkg} shows the the root mean square (RMS) of the Galactic noise trace in the time domain, plotted as a function of the LST in the East-West polarization.

Since the antenna used is omnidirectional, it detects a diffuse background, with no significant correlation between the amplitude and phase of the noise. Therefore, we generated a uniformly distributed phase that was added to noise spectrum. By performing an inverse Fast Fourier Transform (iFFT), we obtained the noise trace in the time domain, which was then added to the voltage trace representing the signal.

Thermal noise in the electronics of the Radio Frequency (RF) chain also contributes to the overall noise. The electronic noise is often added to the signal coming from each antenna arm, that is usually referred to as the open circuit voltage ($V_{\mathrm{oc}}$). However, this work focuses on the reconstruction of the electric field starting from the open circuit voltage, and all the possible noise contributions from the RF chain that could be added between $V_\mathrm{oc}$ and the digitized signal ($V_\mathrm{ADC}$) are not considered. The reconstruction of the open circuit voltage starting from the digitized signal is a topic that will be addressed in future analyses involving real experimental data.

\section{Electric field reconstruction}
 Given that electromagnetic waves in air propagate as transverse waves, the electric field component along the radial direction (aligned with the propagation direction $\hat{r}$) is negligible. This allows for the elimination of one degree of freedom. To simplify the convolution formalism, the voltage is expressed in spherical coordinates as:

\begin{equation}
    \mathcal{V}(f) = \mathcal{V}_{\mathrm{oc}}(f)+\mathcal{N}(f)=
    \begin{pmatrix}
        \mathcal{H}_{1, \theta} & \mathcal{H}_{1, \phi} & \mathcal{H}_{1, r} \\
        \mathcal{H}_{2, \theta} & \mathcal{H}_{2, \phi} & \mathcal{H}_{2, r} \\
        \mathcal{H}_{3, \theta} & \mathcal{H}_{3, \phi} & \mathcal{H}_{3, r}
    \end{pmatrix} \vec{\mathcal{E}}(f)+\mathcal{N}(f)
    \label{eq:v_matrix}
\end{equation}
where $\mathcal{V}_{\mathrm{oc}}$ denotes the open circuit voltage. The subscripts 1,2,3 in the vector $\mathcal{H}$ correspond to the three orthogonal antenna arms, while $\theta$, $\phi$ and $r$ represent the directions of electric field polarization in spherical coordinates \cite{abreu2012antennas}. $\mathcal{N}(f)$ represents the noise added in each polarization.

For this analysis, the galactic noise at LST = 18h is used, as it corresponds to elevated noise levels across all polarizations.

The $r$-component of equivalent length of antenna, as shown in  Figure \ref{fig:ant_respn_dipole} and \ref{fig:ant_resp} is negligible. Consequently, the original 3×3  response matrix can be reduced to a 3×2 formulation as:

\begin{equation}
        \boldsymbol{\mathcal{V}}(f) = 
    \begin{pmatrix}
        \mathcal{H}_{1, \theta} & \mathcal{H}_{1, \phi}\\
        \mathcal{H}_{2, \theta} & \mathcal{H}_{2, \phi} \\
        \mathcal{H}_{3, \theta} & \mathcal{H}_{3, \phi}
    \end{pmatrix}
    \begin{pmatrix}
        \mathcal{E_\theta}\\
        \mathcal{E_\phi}
    \end{pmatrix}+\mathcal{N}(f)
    \label{eq:v_matrix_simp}
\end{equation}

In this study, two electric field reconstruction methodologies are employed:
\begin{itemize}

\item A conventional matrix inversion technique, currently widely adopted in radio detection experiments, which entails inverting the response matrix using either two or three polarizations (section \ref{conv_mthd}). 

\item  An analytical $\chi^2$ minimization method using three polarizations, detailed in (section \ref{least_estm}). This is a novel methodology first introduced in this work.

\end{itemize}

\subsection{Matrix inversion method}
\label{conv_mthd}

When the SNR is sufficiently high across all traces within the operational bandwidth, and the arrival direction is known, two orthogonal horizontal polarizations can achieve accurate electric field reconstruction. This methodology has so far been implemented in most radio detection experiments \cite{abreu2012antennas, schellart2013detecting}. The electric field is reconstructed by inverting Equation \ref{eq:v_matrix_simp} using any two voltage polarizations, expressed as $\vec{\mathcal{E}}=\vec{\mathcal{H}}^{-1} \mathcal{V}$. In this approach, noise is considered to be small enough to assume that $\mathcal{V}_{\mathrm{oc}}(f)$ is equal to  $\mathcal{V}(f)$.

\begin{equation}
    \begin{pmatrix}
        \mathcal{E_\theta}\\
        \mathcal{E_\phi}
    \end{pmatrix}=
    \begin{pmatrix}
        \mathcal{H}_{1, \theta} & \mathcal{H}_{1, \phi}\\
        \mathcal{H}_{2, \theta} & \mathcal{H}_{2, \phi} \\
    \end{pmatrix}^{-1}
        \begin{pmatrix}
        \mathcal{V}_1\\
        \mathcal{V}_2
    \end{pmatrix}
    \label{eq:conv_method2}
\end{equation}

Under noise-free conditions, this method exactly reconstructs the electric field. By incorporating the third polarization, the electric field can also be reconstructed, as first demonstrated in the LOPES-3D experiment \cite{huber2022lopes}.

\begin{figure}[htbp]
\centering
\includegraphics[width=.495\textwidth]{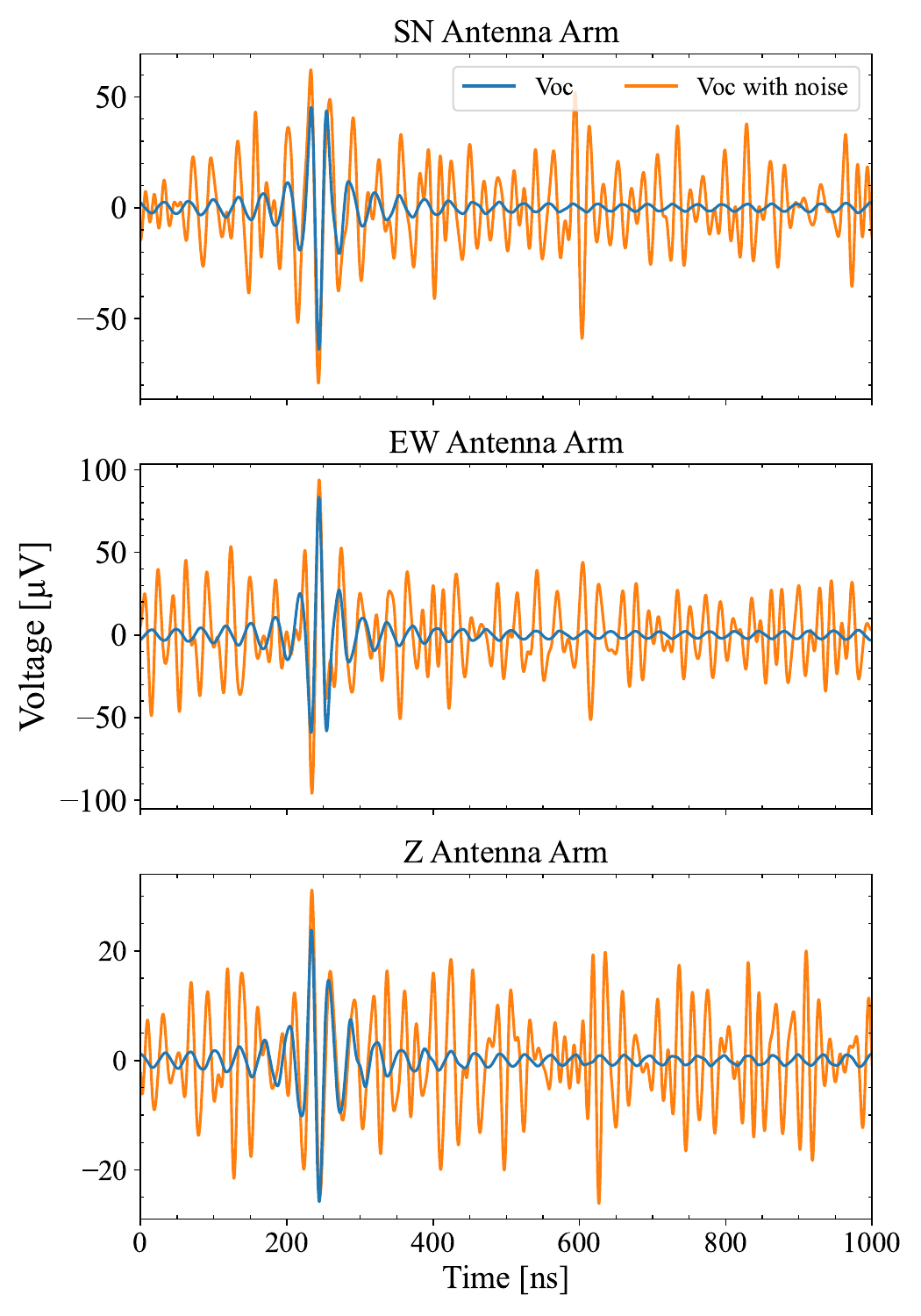}
\includegraphics[width=.495\textwidth]{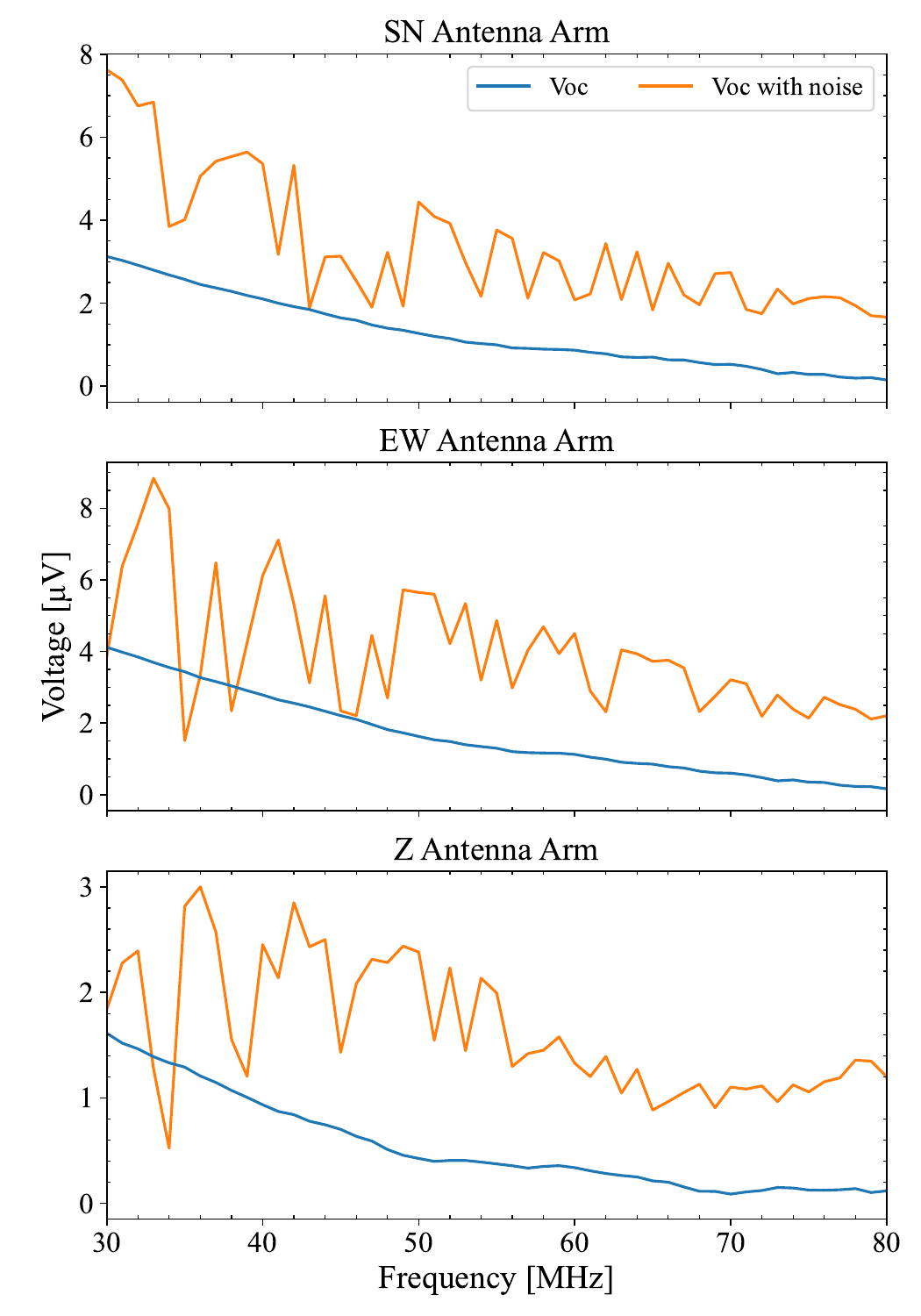}
   \caption{Simulated voltage traces in the time domain (left) and frequency domain (right), generated using the dipole antenna response. The electric field corresponds to  a proton air shower with a primary energy of \unit[0.631]{EeV}, zenith angle $\theta=67.8^\circ$, azimuth angle $\phi=45^\circ$, and a distance of \unit[1009]{m} from the shower core. Simulations incorporate the LFMap galactic noise model.}
   \label{fig:v_trace_dipole_t}
\end{figure}

\subsubsection{Two polarizations vs.\ three polarizations}
\label{2pvs3p}

Although the \(r\)-component is small, the inverse matrix in Equation \ref{eq:v_matrix} can still be found. However, the small magnitude of the \(r\)-component can lead to a significant amplification of noise during the inversion process, potentially causing instabilities or artificial enhancements in the reconstructed signal. This effect becomes particularly critical when the signal-to-noise ratio is low and will be illustrated in the following analysis. Despite this limitation, in general, the electric field can be reconstructed using the three polarizations of the antenna response.

\begin{equation}
    \begin{pmatrix}
        \mathcal{E_\theta}\\
        \mathcal{E_\phi}\\
        \mathcal{E}_{r}
    \end{pmatrix}=
    \begin{pmatrix}
        \mathcal{H}_{1, \theta} & \mathcal{H}_{1, \phi} & \mathcal{H}_{1, r} \\
        \mathcal{H}_{2, \theta} & \mathcal{H}_{2, \phi} & \mathcal{H}_{2, r} \\
        \mathcal{H}_{3, \theta} & \mathcal{H}_{3, \phi} & \mathcal{H}_{3, r}
    \end{pmatrix} ^{-1}
        \begin{pmatrix}
        \mathcal{V}_1\\
        \mathcal{V}_2\\
        \mathcal{V}_3\\
    \end{pmatrix}
    \label{eq:conv_method}
\end{equation}
We compare this approach with the conventional two-polarizations method using a simple dipole model operating in the 30-80 MHz range. The presented example demonstrates voltage traces exhibiting a SNR of approximately 6 in each polarization. Here, the SNR is defined as $S_{\mathrm{peak}}$/RMS$_{\mathrm{noise}}$, where S is the maximum value of the Hilbert envelope of the voltage trace, and the noise root mean square (RMS) of noise corresponds to the standard deviation calculated within a \unit[250]{ns} time window of the voltage trace positioned \unit[500]{ns} after the peak signal. 
Figure \ref{fig:v_trace_dipole_t} illustrates the signals in the three polarizations.



\begin{figure}[htbp]
\centering
\includegraphics[trim={0.5cm 3cm 0.5cm 3cm}, width=0.48\textwidth]{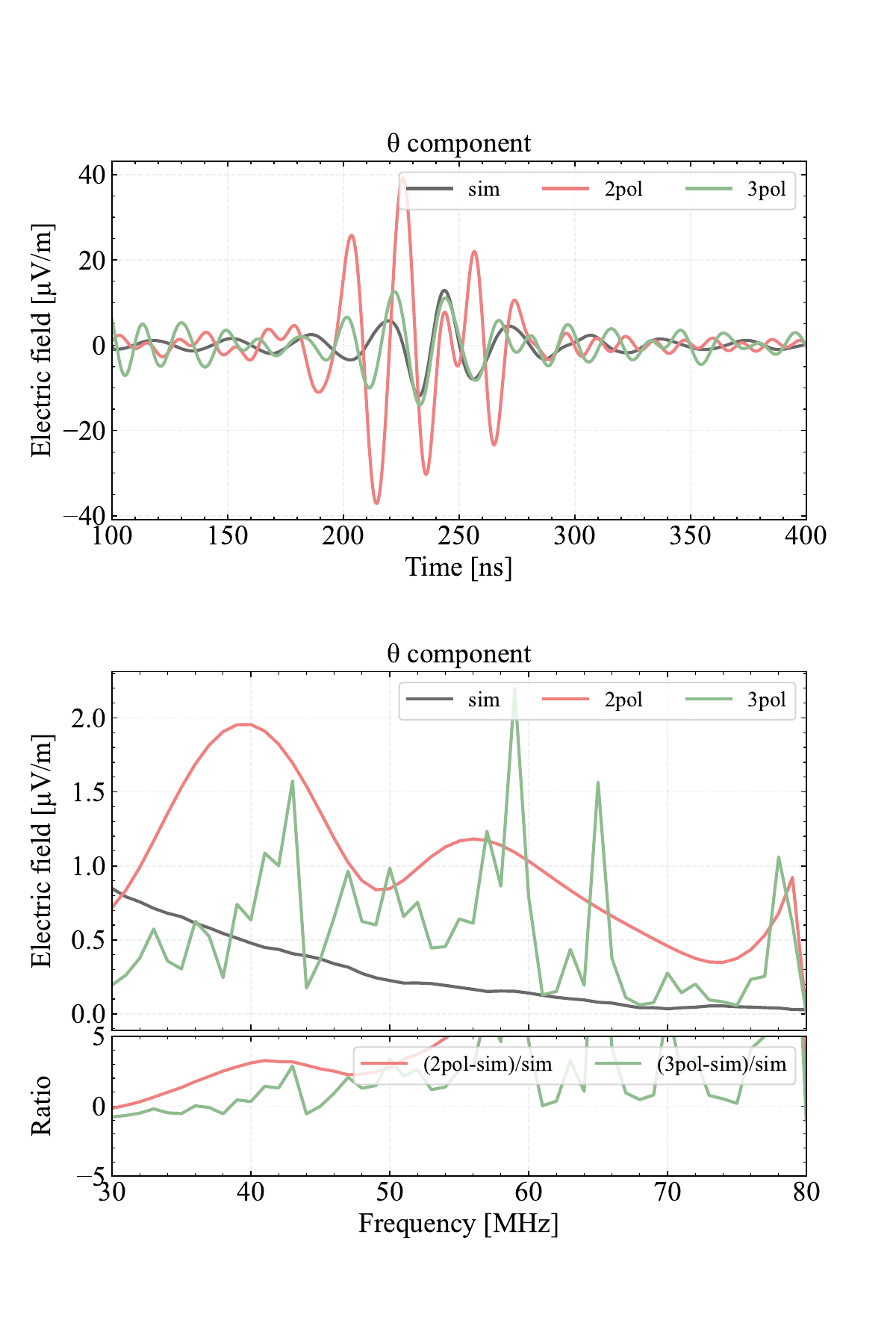}
\includegraphics[trim={0.5cm 3cm 0.5cm 3cm}, width=0.48\textwidth]{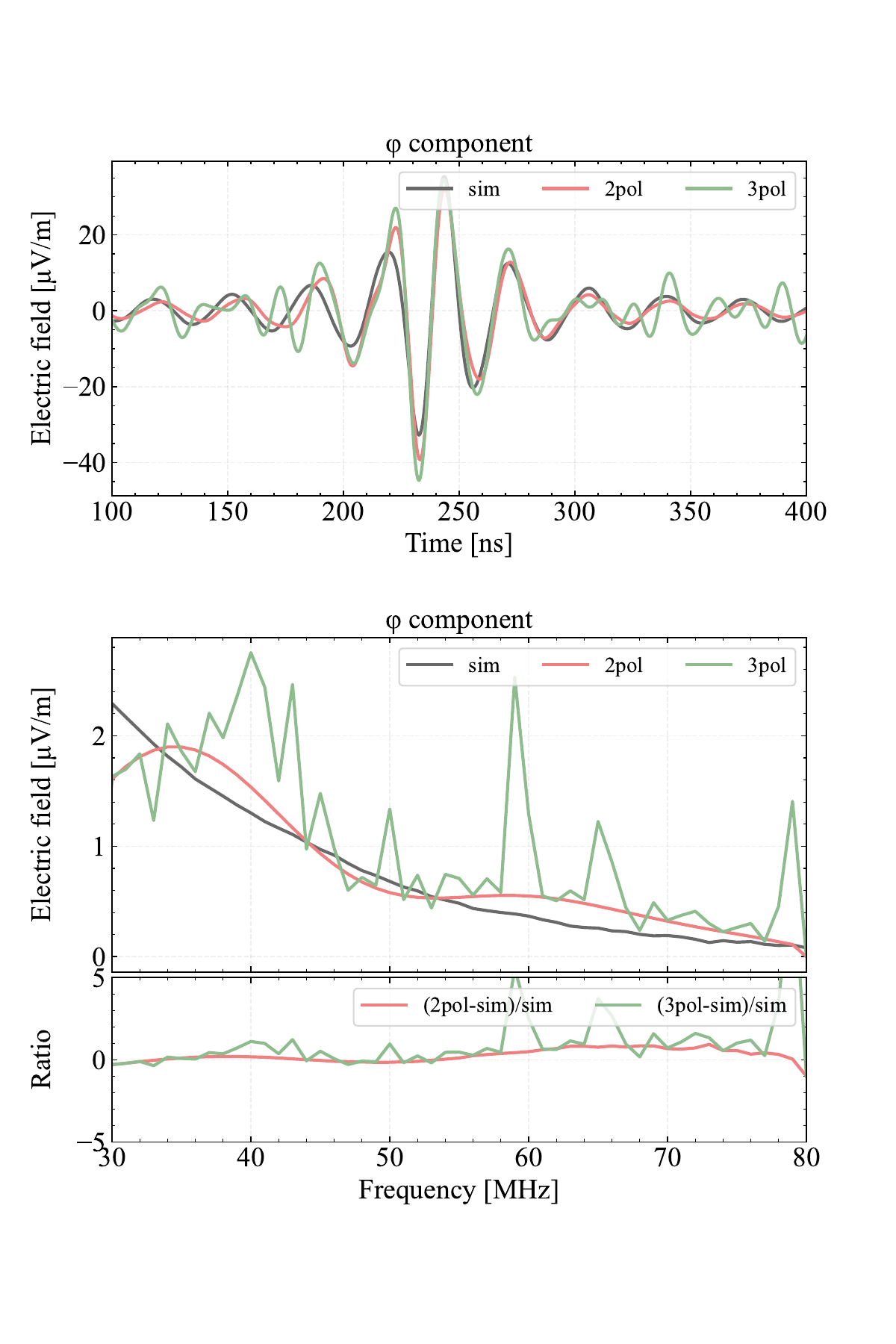}    
\caption{Reconstructed electric field traces derived via the the matrix inversion method from the voltage traces in Figure~\ref{fig:v_trace_dipole_t}, illustrated in both the time domain (top panel) and frequency domain (bottom panel). The colored curves represent the simulation (sim), the matrix inversion method with 2 polarizations (2pol) and 3 polarizations (3pol).}
    \label{fig:e_rec_dipole}
\end{figure}

\begin{center}
\begin{table}[ht]
\centering
\begin{tabular}{lcc}
\hline
\multirow{2}{*}{   Reconstruction} & \multicolumn{2}{c}{Normalized cross-correlation} \\ \cline{2-3}                                 & \multicolumn{1}{c}{$\theta$-component}  & $\varphi$-component \\ 
\hline \hline
2 polarizations (matrix inversion)  & \multicolumn{1}{c}{-0.13}   & 0.93 \\ 
3 polarizations (matrix inversion)  & \multicolumn{1}{c}{0.40}   & 0.80 \\ 
\hline
\end{tabular}
\caption{A comparison of the normalized cross-correlation values for the reconstructed electric field traces using the matrix inversion method with two and three polarizations for the dipole antenna.}
\label{tab:n_cc1}
\end{table}
\end{center}

The reconstruction results for this example are presented in Figure  \ref{fig:e_rec_dipole} and Table \ref{tab:n_cc1}, which offers a quantitative comparison of the quality of the electric field reconstruction using the normalized cross-correlation values between the reconstructed and simulated (Monte Carlo true) electric field traces \cite{corstanje2023high}. 

For the \(\theta\)-component, which exhibits a weaker signal, the three-polarization reconstruction reduces the larger deviations seen in the two-polarization method in the time domain (upper left panel).  However, both of these methods introduce anomalous signal amplifications  presented by the bumps and spikes in the frequency domain (bottom left panel). 
For the $\varphi$-component of the electric field, which exhibits a  stronger signal,  the two-polarization method performs better than the three-polarization method, both in the time domain (upper right panel) and frequency domain (bottom right panel). Especially, in the frequency domain, the three-polarization method introduces anomalous spikes at certain frequencies, similar to that in the $\theta$-component. These artifacts are primarily introduced by the inclusion of the  $r$-component in the reconstruction. This indicates that, although the matrix inversion method offers a solution for three polarizations, further refinement is required to ensure stable and accurate performance.


\subsection{Analytical $\chi^2$ minimization method}
\label{least_estm}

To eliminate the divergences at certain frequencies in the three-polarization reconstruction, we employ a robust $\chi^2$ minimization method. The $\chi ^2$ minimization method is widely utilized across disciplines for model-data optimization and has been previously validated in the forward-folding technique developed for electric field reconstruction in the ARIANNA experiment  \cite{Glaser_2019}. 
High precision results were achieved through iterative $\chi ^2$  minimization, incorporating parameters such as amplitude, frequency slope and phase-offset conditions. In this study, 
we build a single $\chi ^2$ for each antenna with three polarizations, without incorporating any parameters related to the signal characteristics.  Consequently, it offers a more direct and assumption-free framework compared to alternative approaches, enhancing the simplicity of the reconstruction. Given the measured voltage, noise spectrum, and antenna response, the \(\chi^2\) is formulated in frequency domain (denoted in Italian font format) as:

\begin{equation}
\begin{aligned}
    \chi^2 & = \sum_{i=1}^3 \left( \frac{\mathbf{\mathcal{V}}_i - \mathcal{H}_i \begin{pmatrix} \mathcal{E}_\theta \\ \mathcal{E}_\varphi \end{pmatrix} }{\sigma_{\mathcal{V}_i}} \right)^2 \\
    & = (\boldsymbol{\mathcal{V}} - \boldsymbol{\mathcal{H}} \boldsymbol{\mathcal{E}})^T \sigma_{\mathcal{V}}^{-1} (\boldsymbol{\mathcal{V}} - \boldsymbol{\mathcal{H}} \boldsymbol{\mathcal{E}})
\end{aligned}
\end{equation}

where we use the same notation as in Eq. \ref{eq:v_f}. We assume that the error in this measurement is attributed to background noise. Consequently, the covariance matrix is constructed as a diagonal matrix $\sigma_\mathcal{V}=\operatorname{diag}\left(\sigma_{\mathcal{V} 1}, \sigma_{\mathcal{V} 2}, \sigma_{\mathcal{V} 3}\right)$, where the diagonal elements correspond to the squared noise spectrum in each polarization of the antenna respectively.

To minimize $\chi^2$, we derive an analytical solution for the electric field  by equating the gradient of $\chi^2$ to 0, expressed as: 
\begin{equation}
    \nabla_{\mathcal{E}} \chi^2 = -2 (\boldsymbol{\mathcal{H}}^T \sigma_{\mathcal{V}}^{-1} \boldsymbol{\mathcal{V}} - \boldsymbol{\mathcal{H}}^T \sigma_{\mathcal{V}}^{-1} \boldsymbol{\mathcal{H}}\boldsymbol{\mathcal{E}}) = 0
    \label{eq:chi2mini}
\end{equation}
Following the formalism in \cite{Behnke:2013pga}, the reconstructed electric field is expressed as:
\begin{equation}
    \boldsymbol{\mathcal{E}} = (\boldsymbol{\mathcal{H}}^T \sigma_{\mathcal{V}}^{-1} \boldsymbol{\mathcal{H}})^{-1} \boldsymbol{\mathcal{H}}^T \sigma_{\mathcal{V}}^{-1} \boldsymbol{\mathcal{V}}
    \label{eq:e_inverse}
\end{equation}

By performing a simultaneous minimization across all three polarizations, the method captures more information from the electric field. By rigorously accounting for noise contributions, this method ensures further reconstruction accuracy across the full frequency spectrum. 
 

\subsection{Matrix inversion vs. analytical $\chi^2$ minimization}
\label{comparison1}

\begin{figure}[htbp]
\centering
\includegraphics[trim={0.5cm 3cm 0.5cm 3cm}, width=0.48\textwidth]{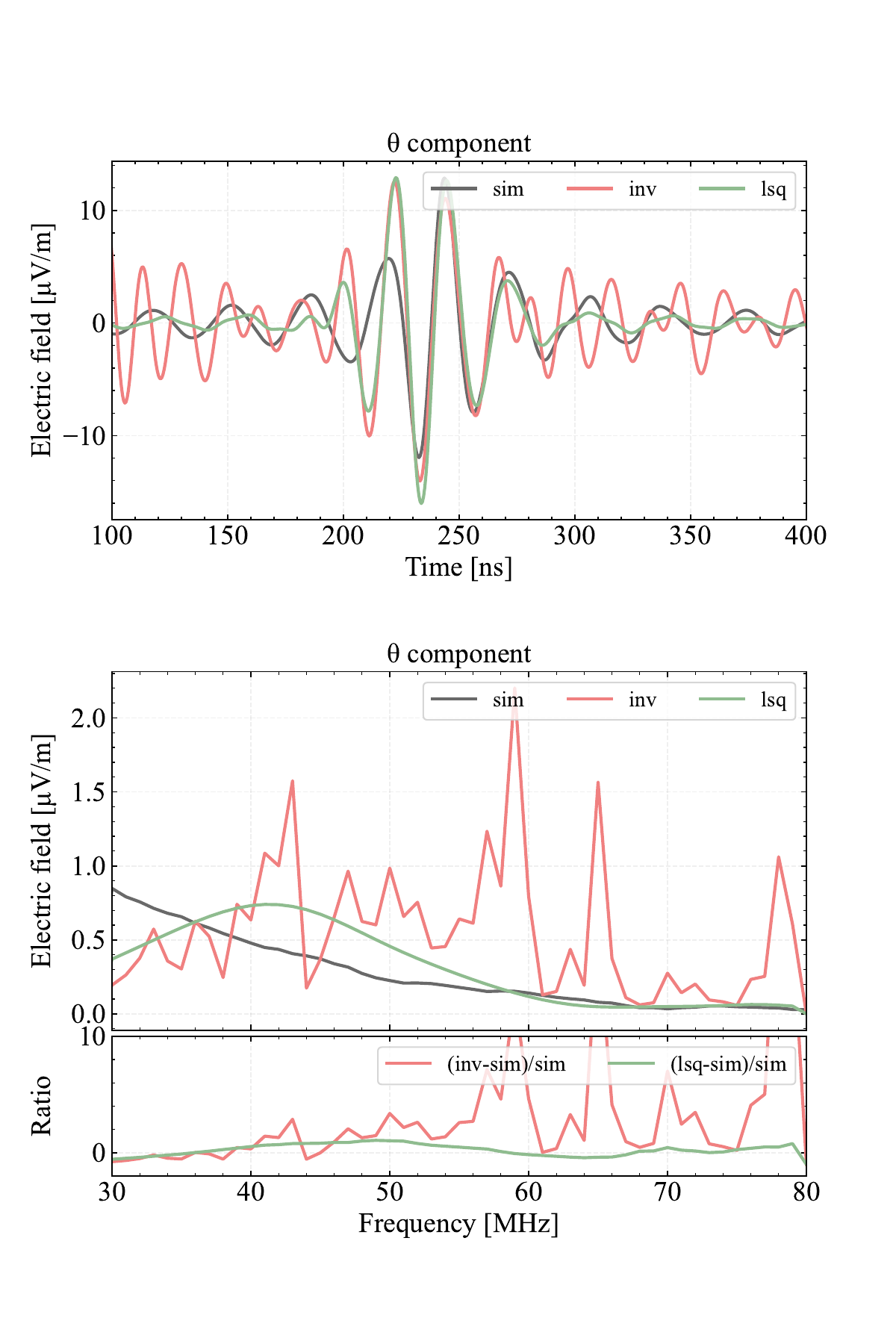}
\includegraphics[trim={0.5cm 3cm 0.5cm 3cm}, width=0.48\textwidth]{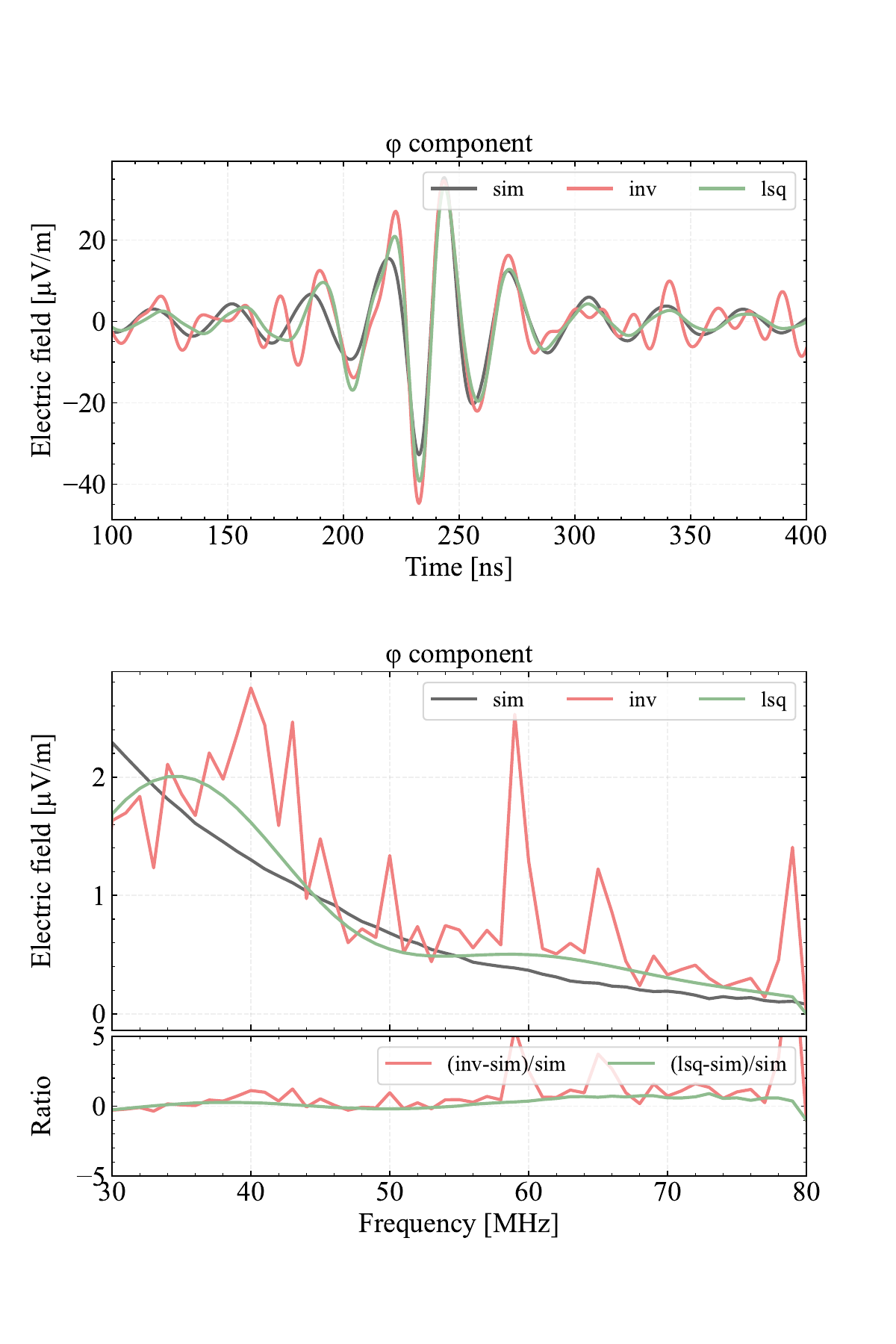}    
    \caption{Comparison of the matrix inversion method (inv) and the analytical $\chi^2$ minimization method (lsq) in the three-polarization reconstruction for the dipole antenna. Top panels: time traces of the electric field. Bottom panels: electric field spectrum in frequency domain. }
    \label{fig:e_rec_dipole_lsq}
\end{figure}

In Figure \ref{fig:e_rec_dipole_lsq}, a comparative analysis is presented between the analytical $\chi^2$ minimization method and the matrix inversion approach for reconstructing the electric field using the dipole antenna.  In the time domain (top panels), for both the $\theta$- and $\varphi$-components, the analytical $\chi^2$ minimization yields better reconstructed traces than those obtained via the matrix inversion approach. The reconstructed time-domain traces exhibit fewer fluctuations within the non-signal time window than those derived from the matrix inversion method. In the frequency domain (bottom panels), for both the $\theta$- and $\varphi$-components, the anomalous spikes generated by the matrix inversion method are flattened in the analytical $\chi^2$ minimization (or least-squares method, denoted as lsq) results. These improvements across both  domains collectively demonstrate the enhanced reconstruction fidelity exhibited by the analytical $\chi^2$ minimization method. 

To quantify  the performance of these two methods, cross-correlation coefficients are computed and tabulated in Table \ref{tab:n_cc2}.
The analytical $\chi^2$ minimization (lsq) yields larger normalized cross-correlation coefficients than those of the matrix inversion method for both components, confirming its superior performance. This example shows how the analytical $\chi^2$ minimization method can effectively minimize noise-induced distortions and ensure more precise results.

\begin{center}
\begin{table}[ht]
\centering
\begin{tabular}{lcc}
\hline
\multirow{2}{*}{Reconstruction} & \multicolumn{2}{c}{normalized cross-correlation} \\ \cline{2-3}                                 & \multicolumn{1}{c}{$\theta$-component}  & $\varphi$-component \\ 
\hline \hline
3 polarizations (matrix inversion)          & \multicolumn{1}{c}{0.40}   & 0.80 \\ 
3 polarizations (lsq)           & \multicolumn{1}{c}{0.77}   & 0.95 \\ 
\hline
\end{tabular}
\caption{Comparison of the normalized cross-correlation values for the electric field traces reconstructed using the matrix inversion method and the $\chi^2$ minimization method (lsq), with three polarizations, using the dipole antenna.}
\label{tab:n_cc2}
\end{table}
\end{center}

\subsection{Application to the HORIZON antenna}

\begin{figure}[htbp]
\centering
\includegraphics[width=.495\textwidth]{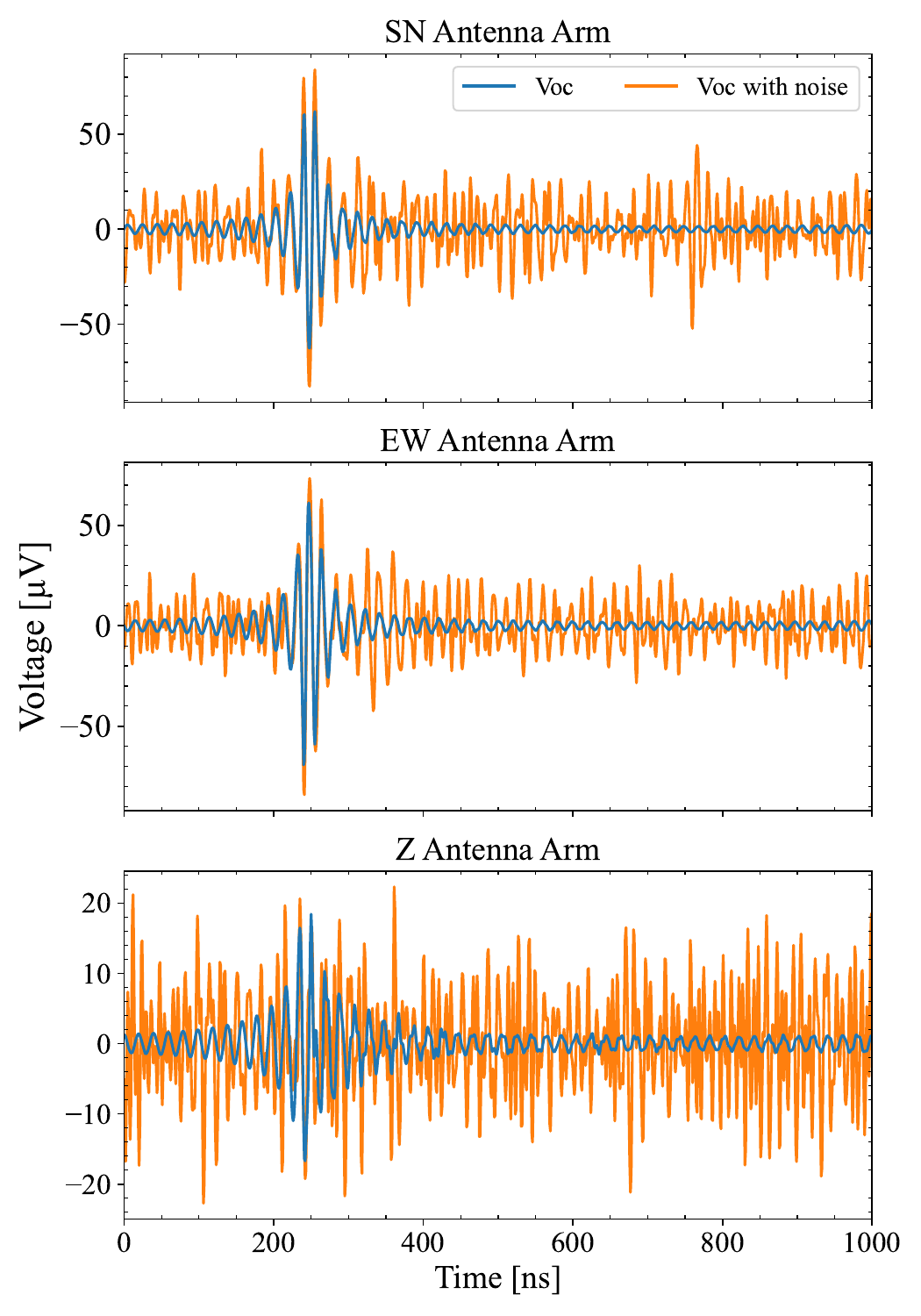}
\includegraphics[width=.495\textwidth]{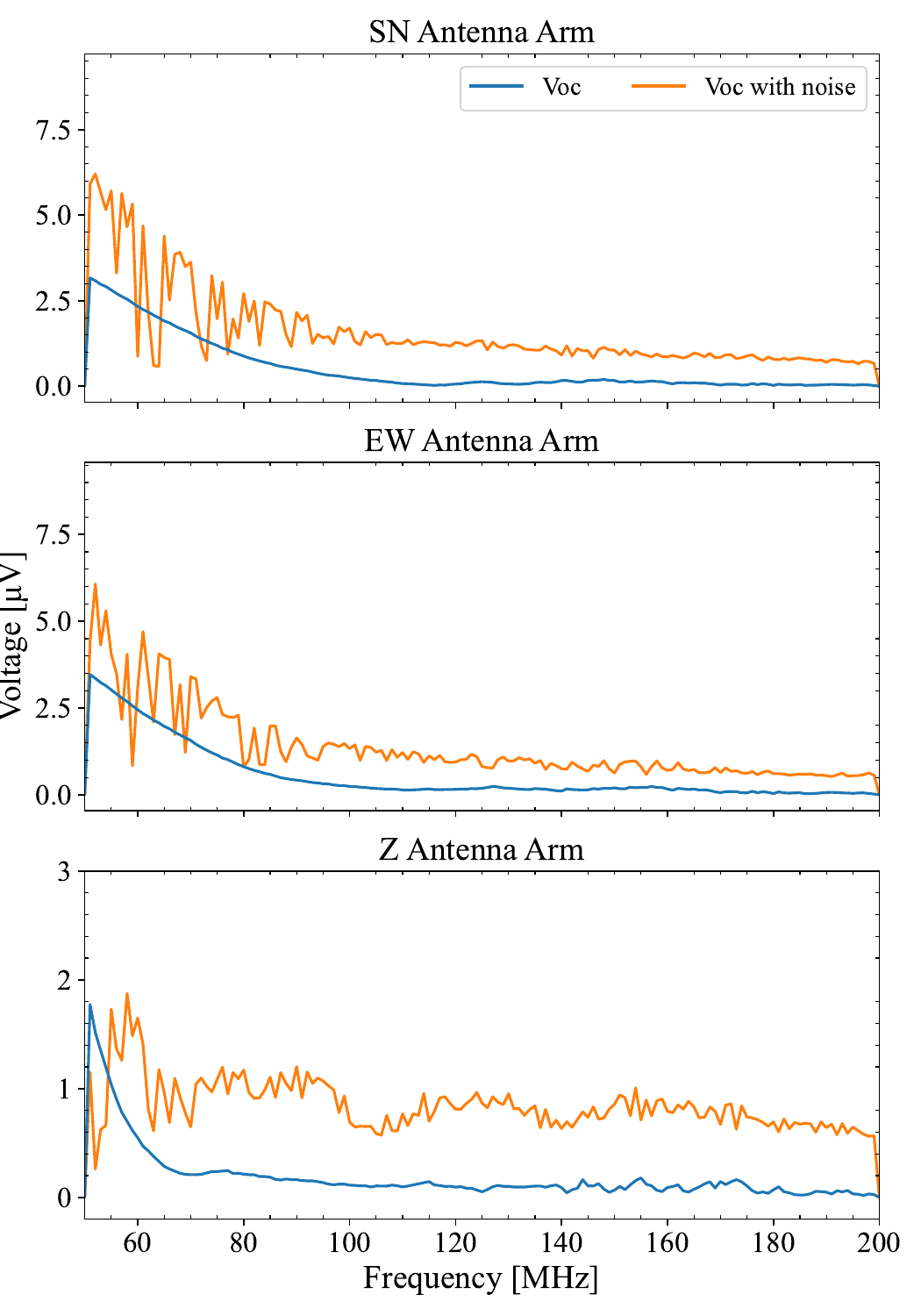}    
    \caption{Simulated voltage traces in the time domain (left) and frequency domain (right), generated  with the HORIZON antenna response. Here,  simulation of air shower and Galactic noise are the same as those for the dipole antenna shown in Figure~\ref{fig:v_trace_dipole_t}, but the plots corresponds to a different trace measured at a distance of \unit[902]{m} from the shower axis.}
    \label{fig:v_trace}
\end{figure}
To verify that the $\chi^2$ method is not dependent on specific antenna responses and frequency bandwidth, we applied it to the HORIZON antenna (see Sec. \ref{sec:Horizon}), which operates with a more complex response over a wider frequency range. Using the same simulation parameters as those applied for the dipole antenna example, we investigated reconstruction discrepancies under similar near-threshold  conditions. Figure \ref{fig:v_trace} shows an example of a voltage trace in both the time and frequency domains.

\begin{figure}[htbp]
\centering
\includegraphics[trim={0.5cm 3cm 0.5cm 3cm}, width=0.48\textwidth]{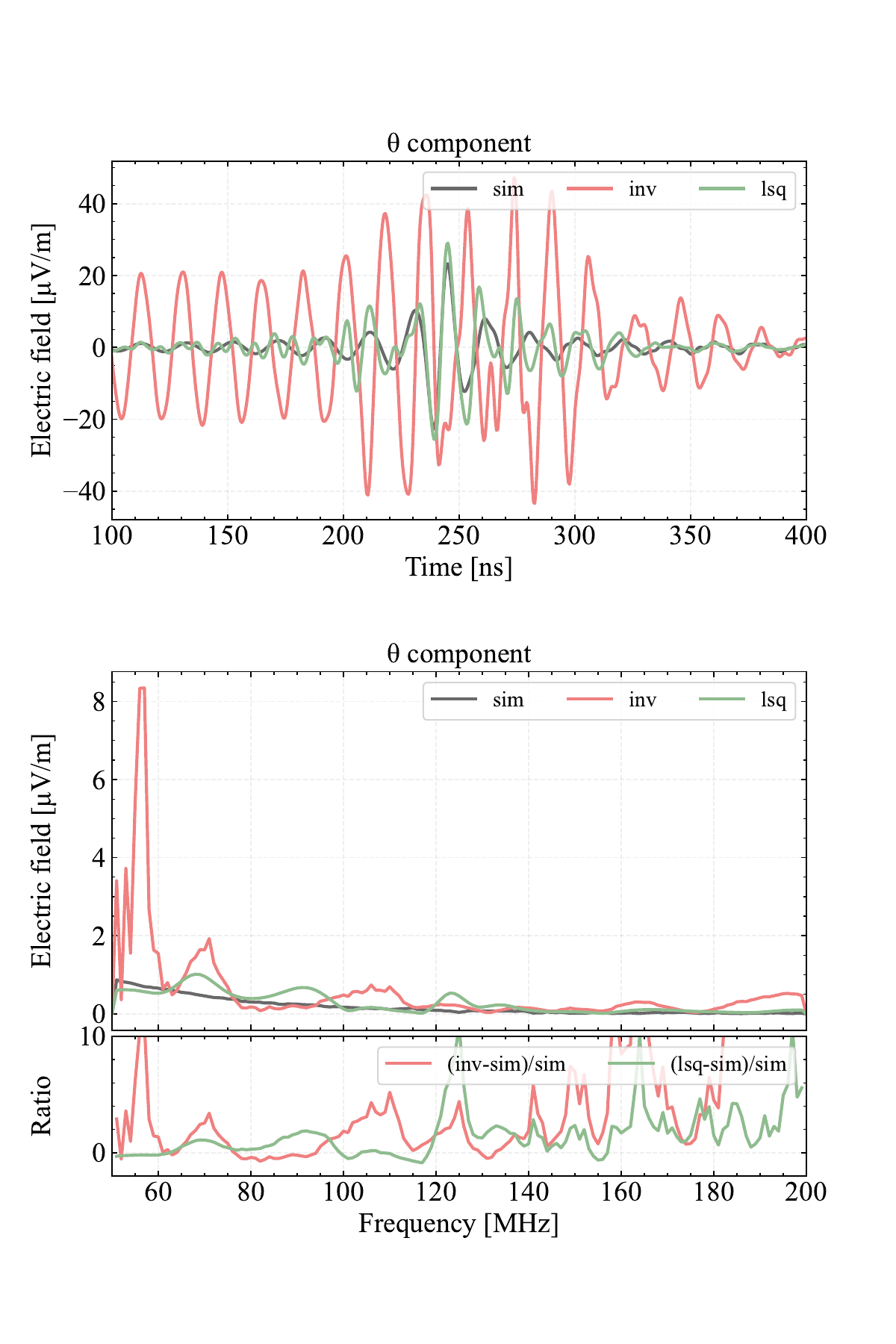}
\includegraphics[trim={0.5cm 3cm 0.5cm 3cm}, width=0.48\textwidth]{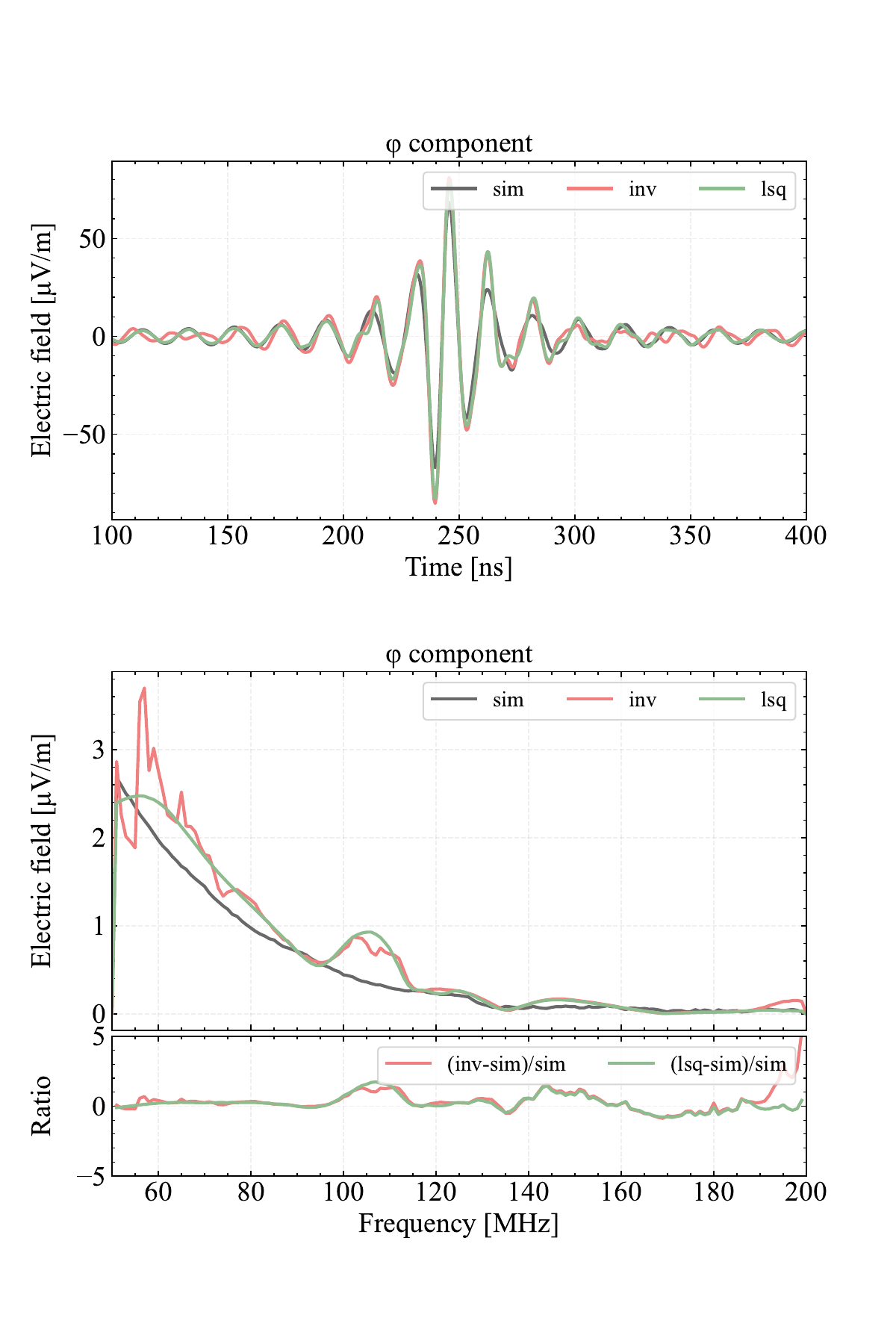}    
    \caption{Comparison of the matrix inversion method (inv) and the analytical $\chi^2$ minimization method (lsq) in the three-polarization reconstruction for the HORIZON antenna. Top panels: time traces of the electric field. Bottom panels: electric field spectrum in the frequency domain. }
    \label{fig:e_rec}
\end{figure}

Figure \ref{fig:e_rec} compares the performance of the matrix inversion and analytical $\chi^2$ minimization on the HORIZON antenna. For the $\varphi$-component (upper right panel), both the analytical $\chi^2$ minimization and matrix inversion methods achieve similar reconstruction within the signal window. 
For the $\theta$-component (upper left panel), characterized by attenuated signal strength, the matrix inversion method fails to reconstruct the electric field trace, exhibiting pronounced anomalous signal distortion. These errors are markedly mitigated by the analytical $\chi^2$ minimization method.

In the frequency domain, for the $\varphi$-component (bottom right panel), there is a bump in the frequency range of approximately \unit[100$-$120]{MHz}, which corresponds to the dip of the $\varphi$-component of the equivalent length within this frequency range, as depicted in Figure \ref{fig:ant_resp}. The equivalent length is smaller below \unit[120]{MHz}, accounting for the increased discrepancies in the reconstructed traces at the lower frequencies generated by the matrix inversion method. These discrepancies are effectively mitigated by the analytical $\chi^2$ minimization.  In the range above \unit[120]{MHz}, the two methods show similar performance, except as the frequency approaches \unit[200]{MHz},  where the equivalent length diminishes and the matrix-inverted trace becomes distorted. For the $\theta$-component (bottom left panel), the superiority of the  analytical $\chi^2$ minimization is more pronounced. The antenna response exhibits a dip at approximately \unit[60]{MHz} in the South-North, East-West arms, and a separate dip near 70 MHz in the vertical arm. Consequently,  significant anomalous amplifications are observed in the matrix-inverted signal around these frequencies. At frequencies above 120 MHz, as the antenna gains decline in the South–North and East–West arms, the matrix-inverted signal exhibits a marked increase in error, as quantified in the error ratio subplot. 

Although both methods exhibit deviations, the analytical $\chi^2$ minimization method provides superior accuracy than the matrix inversion method. This conclusion is further corroborated by the normalized cross-correlation coefficients in Table \ref{tab:n_cc3}. Collectively, the findings presented in this example illustrate the robustness of the analytical $\chi^2$ minimization method against variations in antenna response and signal characteristics.

\begin{center}
\begin{table}[ht]
\centering
\begin{tabular}{lcc}
\hline
\multirow{2}{*}{Reconstruction} & \multicolumn{2}{c}{normalized cross-correlation} \\ \cline{2-3} 
                                & \multicolumn{1}{c}{$\theta$-component}  & $\varphi$-component \\ 
\hline \hline
3 polarizations (inv)          & \multicolumn{1}{c}{-0.34}   & 0.96 \\ 
3 polarizations (lsq)           & \multicolumn{1}{c}{0.79}   & 0.97 \\ \hline
\end{tabular}
\caption{Comparison of the normalized cross-correlation values for the electric field traces reconstructed using the matrix inversion method and the analytical $\chi^2$ minimization method (lsq), with three polarizations, using the HORIZON antenna.
}
\label{tab:n_cc3}
\end{table}
\end{center}

\section{Statistical performance of the analytical $\chi^2$ minimization method}
We now estimate the resolution of the analytical $\chi^2$ minimization method by analyzing a large number of simulated traces described in section \ref{sim_set}. Both the dipole and HORIZON antenna models, as introduced in section \ref{ant_resp}, have been analyzed.

\subsection{Event selection}
Before the statistical analysis, we must select the events that can be  distinguished from the background noise. The selection criteria for identifying simulated signals are informed by the characteristics of the simulated voltage and the background noise. The following criteria are intended to ensure that the detected signals correspond to a shower event:

\begin{itemize}
    \item SNR $>$ 5 in at least one antenna arm.
    \item $t_{\text{signal}} - \unit[200]{ns}$ < $t_{\text{peak}}$<\ \ $t_{\text{signal}}$ + \unit[200]{ns}   : the time of the maximum amplitude of the trace, $t_{\text{peak}}$, must fall within $\pm$\unit[200]{ns} ns of the time the signal is expected to be at in the trace.
    \item Only the innermost 16 antennas out of the 20 simulated antennas in each arm of the star-shaped layout are used. The simulation footprint covers twice the Cherenkov angle, and antennas beyond ring number 16 are excluded due to their negligible signal contribution.
    \item $N_{\text{antenna}} \ \ \geqslant$ 5 : at least 5 antennas must meet the above criteria.
\end{itemize}

For the dipole antenna, 2332 out of 4160  simulations satisfy  these event-selection criteria.  In these selected events, 79\% of the antennas were triggered. For the HORIZON antenna, 2749 simulations passed the selection criteria, with 71\% of antennas triggered accordingly. 

\subsection{Comparison of the peak envelope amplitudes}
\label{hilbert_cmp}
\begin{figure}[!htbp]
\centering
    \includegraphics[width=1.00\textwidth]{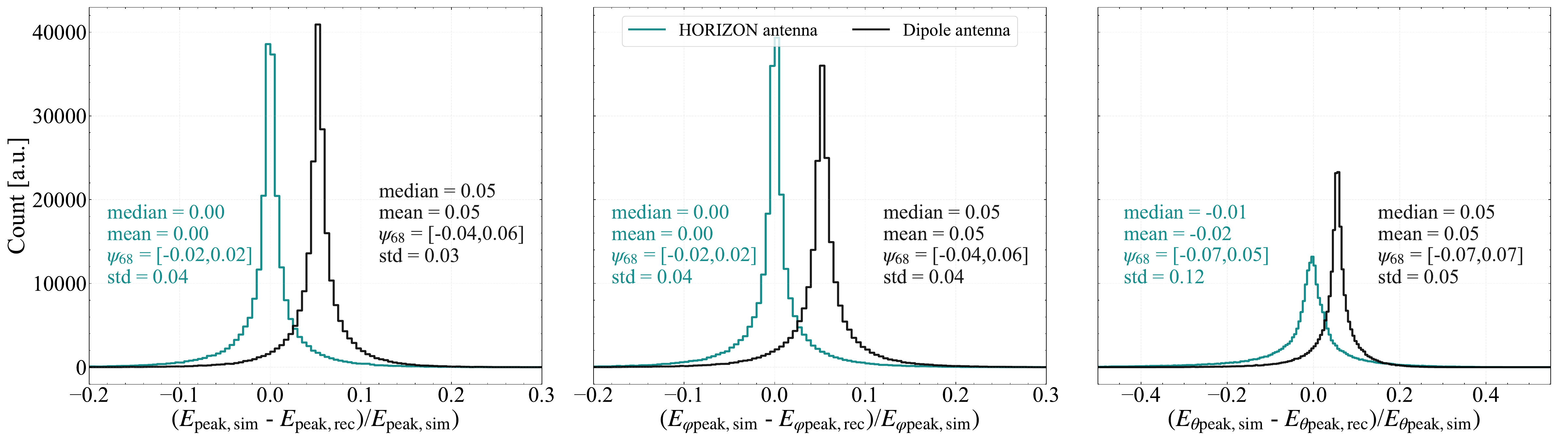}
    \caption{The relative error distribution of the reconstructed PEA in comparison with the Monte Carlo true PEA, depicted across three panels: total(left), $\varphi$-component (middle), and $\theta$-component (right). Results are shown for the HORIZON antenna (green) and the dipole antenna (black). Corresponding statistical metrics are displayed in matching colors.
    } 
    \label{fig:e_comp}
\end{figure}

The Hilbert envelope offers a smooth, amplitude-tracking representation of the signal, rendering it particularly effective for processing oscillatory or noise-contaminated data. To quantify discrepancies, the peak envelope amplitude (PEA)  derived from the electric field is employed to compare the simulated and reconstructed traces. The relative error distribution for this comparison is illustrated in Figure \ref{fig:e_comp}.

The distribution demonstrates that the reconstruction method exhibits robust performance for both antenna types, with the relative error distributions centered approximately around zero, indicating high accuracy in reconstructing the electric field. However, small statistical discrepancies in the mean, median, confidence interval, and standard deviation (std) are observed between the dipole and HORIZON antennas.

For both antennas, the $\varphi$-component exhibits consistently higher reconstruction precision relative to the $\theta$-component. This advantage arises from superior antenna gain across most of the frequency band, as shown in  Figure \ref{fig:ant_respn_dipole} and Figure \ref{fig:ant_resp}, and stronger signal amplitudes in the horizontal arms, as shown in Figure \ref{fig:v_trace_dipole_t} and Figure \ref{fig:v_trace}. The disparity is particularly pronounced for the HORIZON antenna (right panel of Figure \ref{fig:e_comp}), which exhibits a larger standard deviation and a narrower high-gain operational range in the $\theta$-component over its broader frequency range.

For the dipole antenna, the results show a std of 0.04 in the relative error distribution, with a mean and median bias of 0.05, indicating a slight systematic underestimation in the reconstruction values (left panel of Figure \ref{fig:e_comp}). This bias is further underscored by a mild asymmetry in the confidence interval ($\psi_{68}$) bounds.

For the HORIZON antenna, the reconstruction exhibits a reduced systematic bias, with a mean and median of zero, demonstrating excellent fidelity to the true signal. However, for the HORIZON antenna we also observe a marginally broader error distribution, characterized by a std of 0.04. This broadening is most pronounced in the $\theta$-component, where the standard deviation increases to 0.12. Notably, this effect primarily arises from a limited subset of traces exhibiting significant reconstruction errors. These occasional errors stem from the fact that there are approximately 5\% more selected traces than in the dipole case, which incorporates events spanning a wider range of SNRs.  While the HORIZON antenna’s extended frequency bandwidth enhances signal reception, it simultaneously introduces greater variation in antenna response. Consequently, poor signal reconstruction occurs in a minority of cases, resulting in infrequent yet pronounced deviations in reconstruction accuracy. Nevertheless, in general, these findings validate that the analytical $\chi^2$ minimization method achieves a robust reconstruction performance even with the HORIZON antenna.


\subsubsection{Two polarizations vs. three polarizations} 
\label{comparison2v3_lsq}

\begin{figure}[htbp]
\centering
\includegraphics[width=0.55\textwidth]{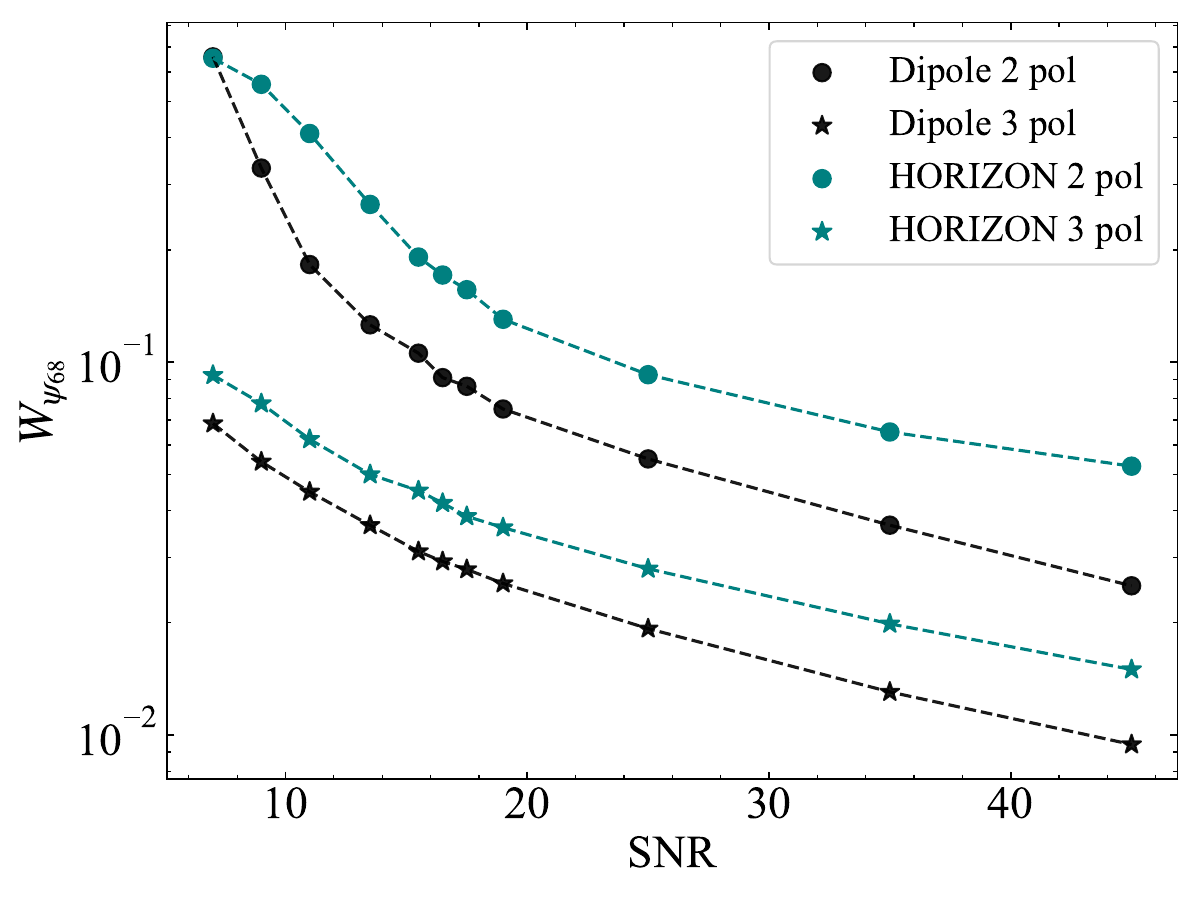}
    \caption{$W_{\psi_{68}}$ of the relative error distribution of the  PEA as a function of SNR using two polarizations (dot) and three polarizations (star) for the dipole antenna (black) and the {\sc HORIZON} antenna (green).}
    \label{fig:snr}
\end{figure}
The addition of the vertical polarization improves the precision of the electric field reconstruction using the analytical $\chi^2$ minimization method.
To objectively compare the resolution of results obtained using two- and three-polarization configurations independent of systematic bias, we employ the half-width of the  68\% confidence interval $W_{\psi_{68}}$  – defined as 34th percentile above and below the median of the relative error distribution, to compare the resolution of the results obtained using 2 and 3 polarizations independent of bias. Figure \ref{fig:snr} illustrates the $W_{\psi_{68}}$  associated with the PEA plotted as a function of SNR.

Adding the vertical polarization improves the reconstruction accuracy across the full SNR range by a factor of 3 to 5, or even more at low SNR, underscoring the critical role of the vertical polarization in refining the reconstruction of the electric field.

\subsubsection{Dependence on the arrival direction of signal}

\begin{figure}[htbp]
\centering
\hspace*{15pt}
\includegraphics[width=1.0\textwidth]{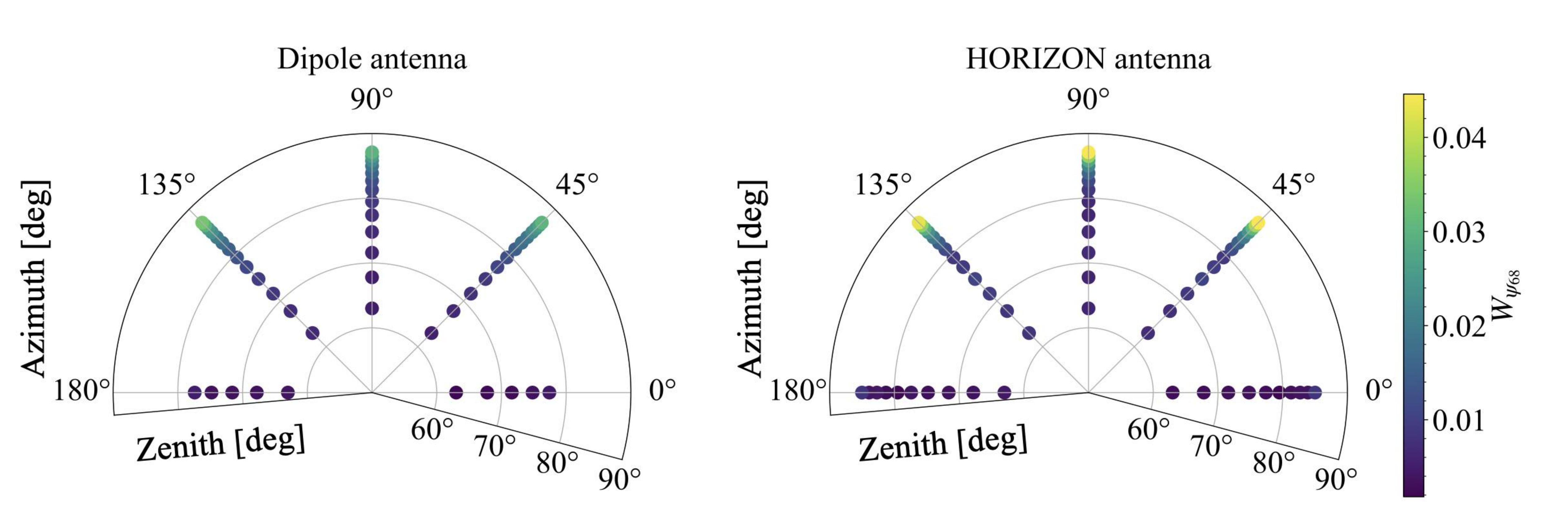}
    \caption{$W_{\psi_{68}}$ of the relative error distribution of the PEA of the reconstructed electric field with respect to the arrival direction for the dipole antenna (left) and the HORIZON antenna (right).}
    \label{fig:dir_cmp}
\end{figure}

As the electric field is convolved with the antenna response, which depends on the arrival direction of the air shower, the resolution of our method  exhibits inherent direction-related uncertainties. Different types of antennas are sensitive to different incident angles, leading to natural, albeit small, variations in precision. This directional dependency is illustrated in Figure \ref{fig:dir_cmp}.

For the PEA of the reconstructed electric field in both antenna configurations, $W_{\psi_{68}}$ of the relative error distribution varies symmetrically with respect to the azimuth angle of $\pi$/2. For the range from 0 to $\pi$, it first rises and subsequently declines, mirroring the symmetric nature of the antenna responses in both the  $\theta$- and $\varphi$-polarizations, as shown in Figure \ref{fig:response_60MHz} of appendix \ref{Responses}. Specifically, in the $\varphi$-component, which dominates the electric field amplitude,  the equivalent length increases when the azimuth angle is smaller than $\pi$/2 and decreases when the azimuth angle is larger than $\pi$/2,  despite the  opposite behavior observed in the $\theta$-component.

For the dipole antenna, precision {presented by $W_{\psi_{68}}$} decreases as the zenith angle increases, which is primarily due to the reduced signal strength at larger zenith angles, even though the equivalent length of the dipole antenna shows  a small increase with increasing zenith angle (see left panel of Figure \ref{fig:response_60MHz} in Appendix \ref{Responses}). In contrast, for the HORIZON antenna, the precision does not follow a consistent trend with zenith angle, there is a small dip around 80$^\circ$  (see left panel of Figure \ref{fig:dir_dep_fix} in Appendix \ref{dir_fix}).  This is because the antenna response, is non-monotonous across the full zenith-angle range (see right panel of Figure \ref{fig:response_60MHz} in Appendix \ref{Responses}). Consequently, the complex variations in precision are influenced by both the irregular antenna response and the changes in signal strength.

As shown in Figure \ref{fig:dir_cmp}, we observe that for the dipole antenna, no event is selected at azimuth angle = 0 or $180^\circ$ and zenith angle  $> 80^\circ$, this is related to the characteristics of the antenna response.  This lack of selected events has not been found with the HORIZON antenna. Despite its larger standard deviation for azimuth angles in the range [45$^\circ$, 135$^\circ$] when zenith angle $>$ 85$^\circ$, the broader frequency band of the HORIZON antenna allows for more signals with amplitudes close to the threshold to be captured, which results in more selected events, making it suitable for detecting inclined air showers. However, as the complex shape of the HORIZON antenna leads to a strong decrease of its equivalent length in certain ranges of arrival directions (see the right panels of Figure \ref{fig:response_60MHz} and \ref{fig:response_60MHz_p_com} in Appendix \ref{Responses}), this will produce larger deviations if the incorrect arrival direction is used.

In conclusion, the analytical $\chi^2$ minimization method shows robust performance in reconstructing the electric field for inclined air showers across a wide range of arrival directions.

\subsubsection{Misreconstruction of the arrival direction of signal}
\label{appd:dir_mis}

In offline analysis, sub-degree precision in direction reconstruction is traditionally attained by analyzing the spatial distribution of electric field arrival times on the ground \cite{Apel_2014, Corstanje_2015, decoene2020sources}. However, an accurate reconstruction of the electric field depends on the antenna response,  which necessitates prior knowledge of the arrival direction as input. Although further refinements could incorporate an improvement of the smoothness of the antenna response, or an iterative direction reconstruction, this study introduces  a straightforward analysis to demonstrate that a directional deviation on the order of one degree exerts negligible influence on the electric field reconstruction. This robustness is largely due to the smooth variation of the antenna gain pattern, that even in the case of the HORIZON antenna allows the target precision to be maintained.

We introduced artificial perturbations to the shower direction, deviating it from the true direction according to a Gaussian distribution with a standard deviation of $1^\circ$ degree in both zenith and azimuth angles. This magnitude of deviation is comparable to the Cherenkov angle in air and is larger than the angular resolution achieved by direction reconstruction methods utilizing timing information and wavefront analysis \cite{Corstanje_2015}.

\begin{figure}[htbp]
\centering
    \includegraphics[width=0.5\textwidth]{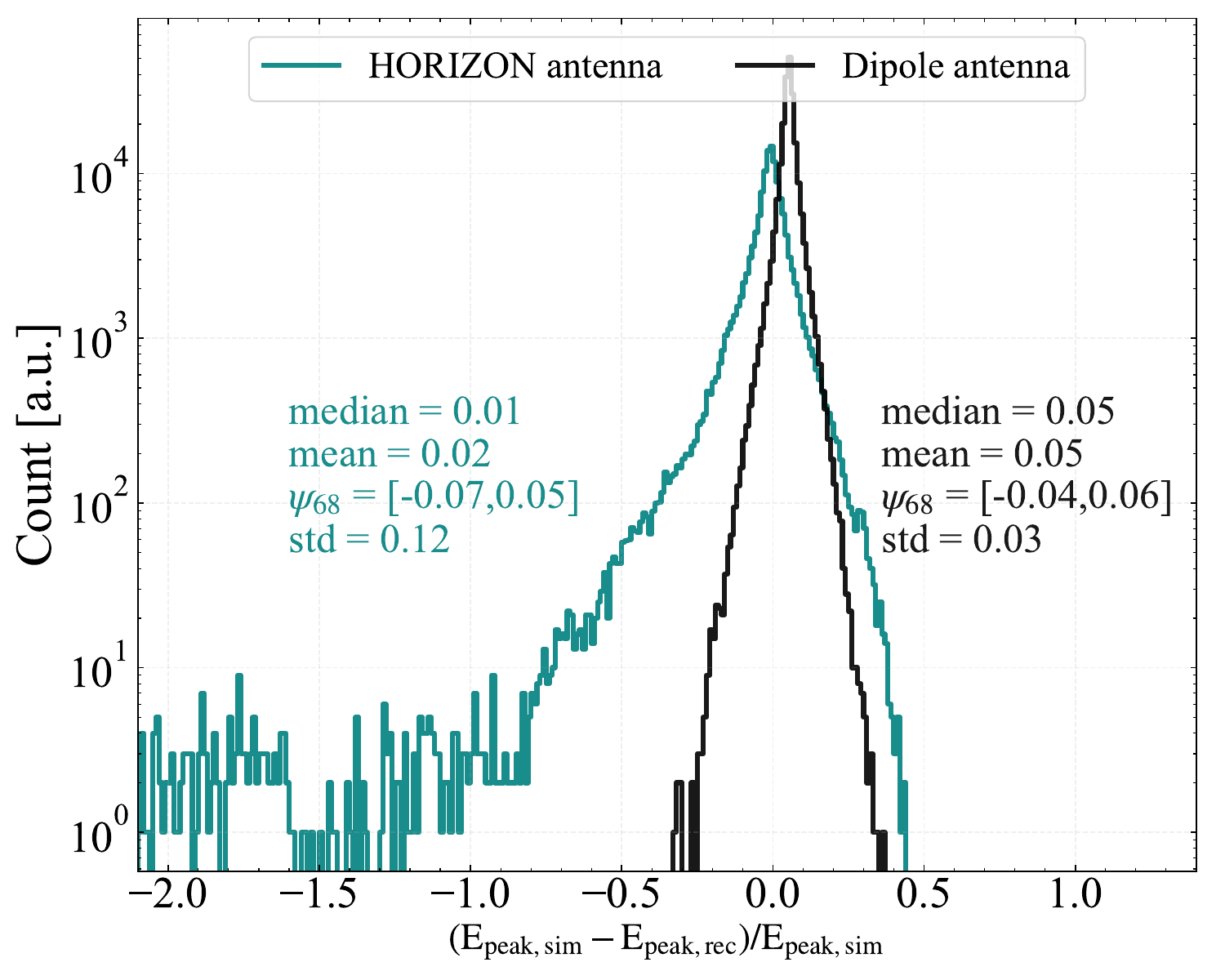}
  \caption{Relative error distribution of the PEA of the reconstructed electric field. The dipole (black) and HORIZON (green) antennas are presented, respectively.  A logarithmic scale is used to clearly show the deviated portion. }
  \label{fig:dir_smearing}
\end{figure}

\begin{figure}[htbp]
\hspace{0.5em}
\centering
    \includegraphics[trim={90pt 0 0pt 0}, width=1.1\textwidth]{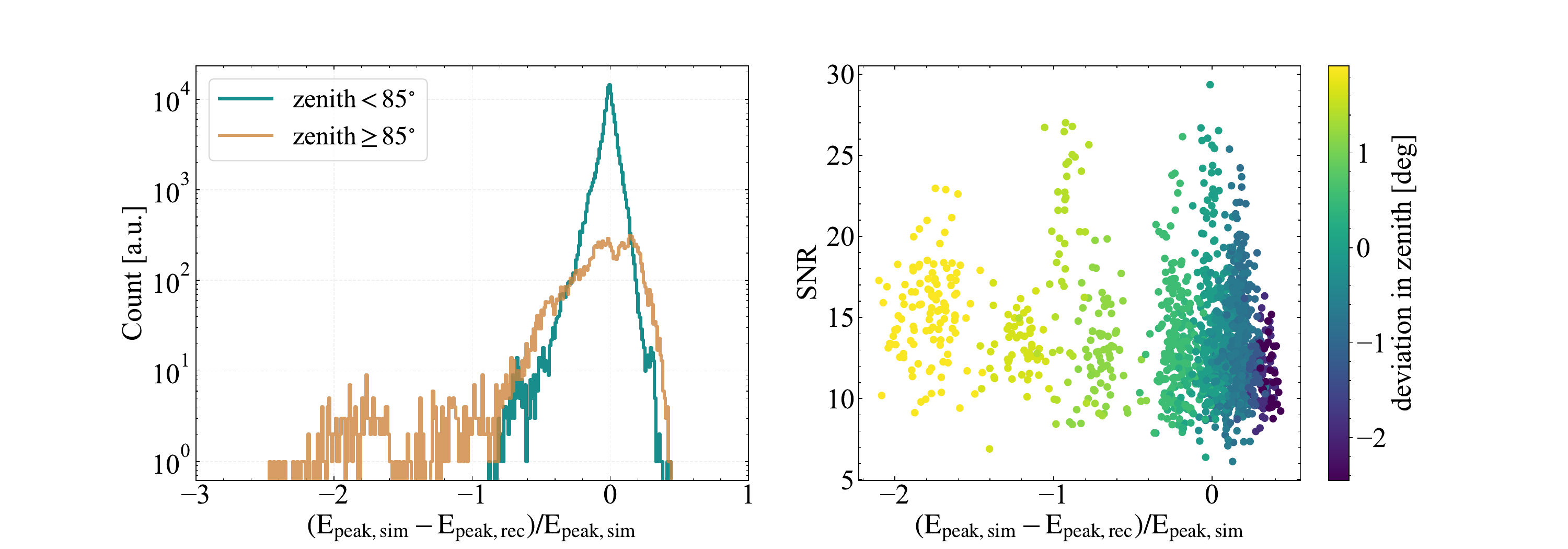}
  \caption{Left panel: Relative error distribution of the PEA of the reconstructed electric field for the HORIZON antenna, for zenith angle = $85^\circ$. Right panel: SNR as a function of relative error for zenith angle $=87.1^\circ$, color-coded by deviation in zenith angle.}
  \label{fig:dir_mis_g}
\end{figure}

Figure \ref{fig:dir_smearing} displays the relative error distribution of the PEA for the reconstructed electric field for both the dipole and HORIZON antennas. Whereas the distribution of  the dipole antenna exhibits minimal variation,  the HORIZON antenna’s results reveal an extended tail characterized by systematic overestimation of the PEA, signifying a reduction in accuracy. A detailed analysis of this phenomenon for the HORIZON antenna is provided in Figure \ref{fig:dir_mis_g}. The left panel separates the relative error distribution of the HORIZON antenna by zenith angle (events below vs. above  $85^\circ$), demonstrating that deviations predominantly cluster at larger zenith angles.  The right panel investigates the dependency of these errors on SNR and zenith angle deviations.   For electric field reconstructions  with larger error, no strong correlation with SNR is observed.  However, relative errors exceeding 1 are primarily associated with zenith angle deviations greater than 1$^\circ$.  This asymmetry originates from directional deviations at specific zenith angles, where there is prominent structure in the antenna gain pattern on the scale of 1$^\circ$. In such cases, the response deviates significantly from the correct one and results in poor reconstructions — a consequence of the antenna’s non-smooth response over frequency and arrival direction.

\subsection{Comparison of energy fluence}
The electromagnetic component of an air shower carries the majority of the primary particle’s energy  and can be estimated from the radiation energy \cite{Glaser:2016qso}. At each antenna, the electric field can be used to determine the energy fluence, which is the radiation energy deposited per unit area.  A spatial integral of the energy fluence across the antenna array yields the total radiation energy within the detector’s frequency band. The energy fluence is given by \cite{Aab_2016, welling2021reconstructing}:


\begin{equation}
    \Phi = c \cdot \epsilon_0 \cdot  \int \vec{E}^2(t) dt
    \label{eq:energy_f}
\end{equation}
with \( \vec{E}\) as the reconstructed electric field, c the speed of light in vacuum and $\epsilon_0$ the vacuum permittivity.
\begin{figure}[htbp]
\hspace{2em}
\centering
\includegraphics[width=1.03\textwidth]{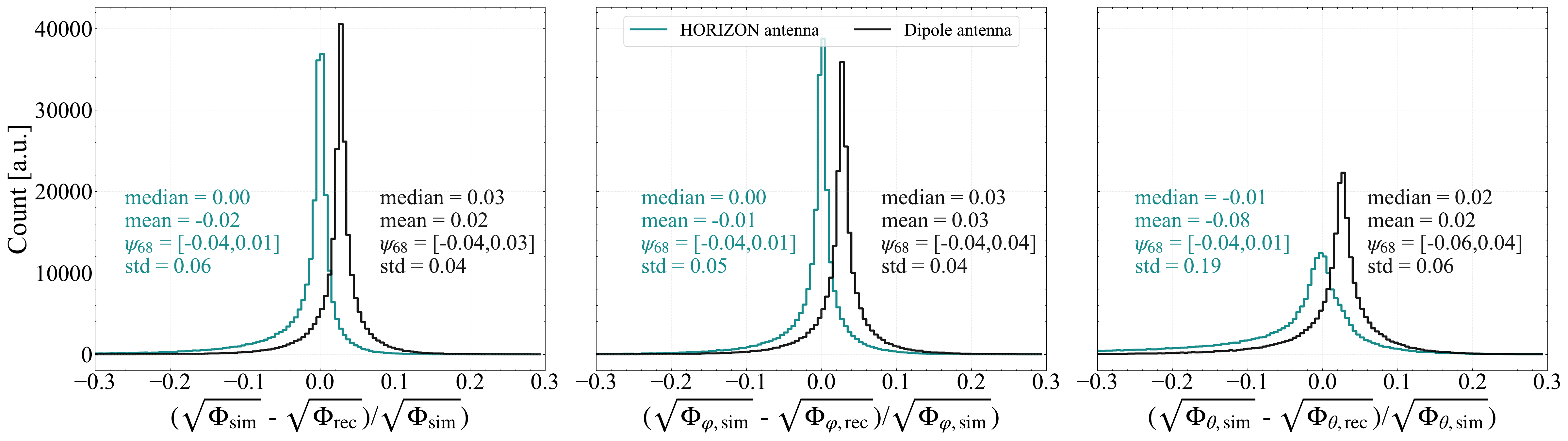}
    \caption{Distribution of relative error in energy fluence. Total 
 (left), $\varphi$ (middle), and $\theta$ (right) components for the HORIZON antenna (green) and the dipole antenna (black). }
    \label{fig:fluence_cmp}
\end{figure}

Radio emission carries information from all stages of air shower development. The typical duration of radio signal ranges from a few nanoseconds to several hundred nanoseconds, with the majority of radiation energy  concentrated in an initial sharp pulse lasting around 10 nanoseconds, followed by a subsequent lower-amplitude and long-duration tail. To estimate energy fluence, two time windows were selected: a \unit[100]{ns} time window symmetrically centered around the signal peak for the signal fluence; a \unit[100]{ns} window at the trace terminus for the noise  fluence quantification. The signal energy fluence was computed by subtracting the background noise fluence from the signal window fluence.

Figure \ref{fig:fluence_cmp} illustrates the relative error distribution in energy fluence, which exhibits a similar trend to the PEA results but exhibiting amplified discrepancies due to error propagation in the squared electric field integration. This enhancement in deviation stems from error accumulation during integration, which cannot be fully suppressed.

\begin{figure}[htbp]
\centering

\hspace{0.6em}
\includegraphics[width=0.48\textwidth]{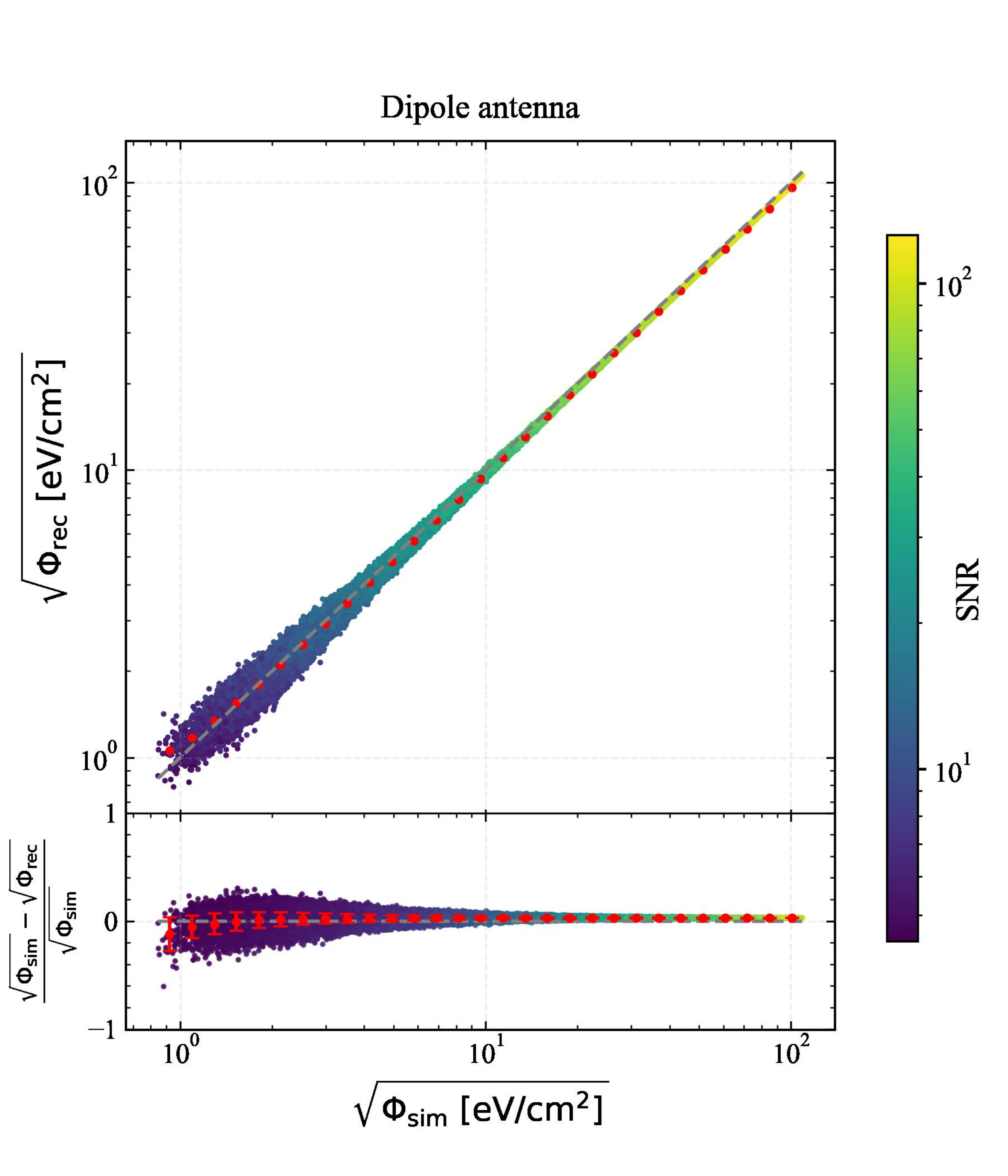}
\hspace{0.1em} 
\includegraphics[clip,width=0.48\textwidth]{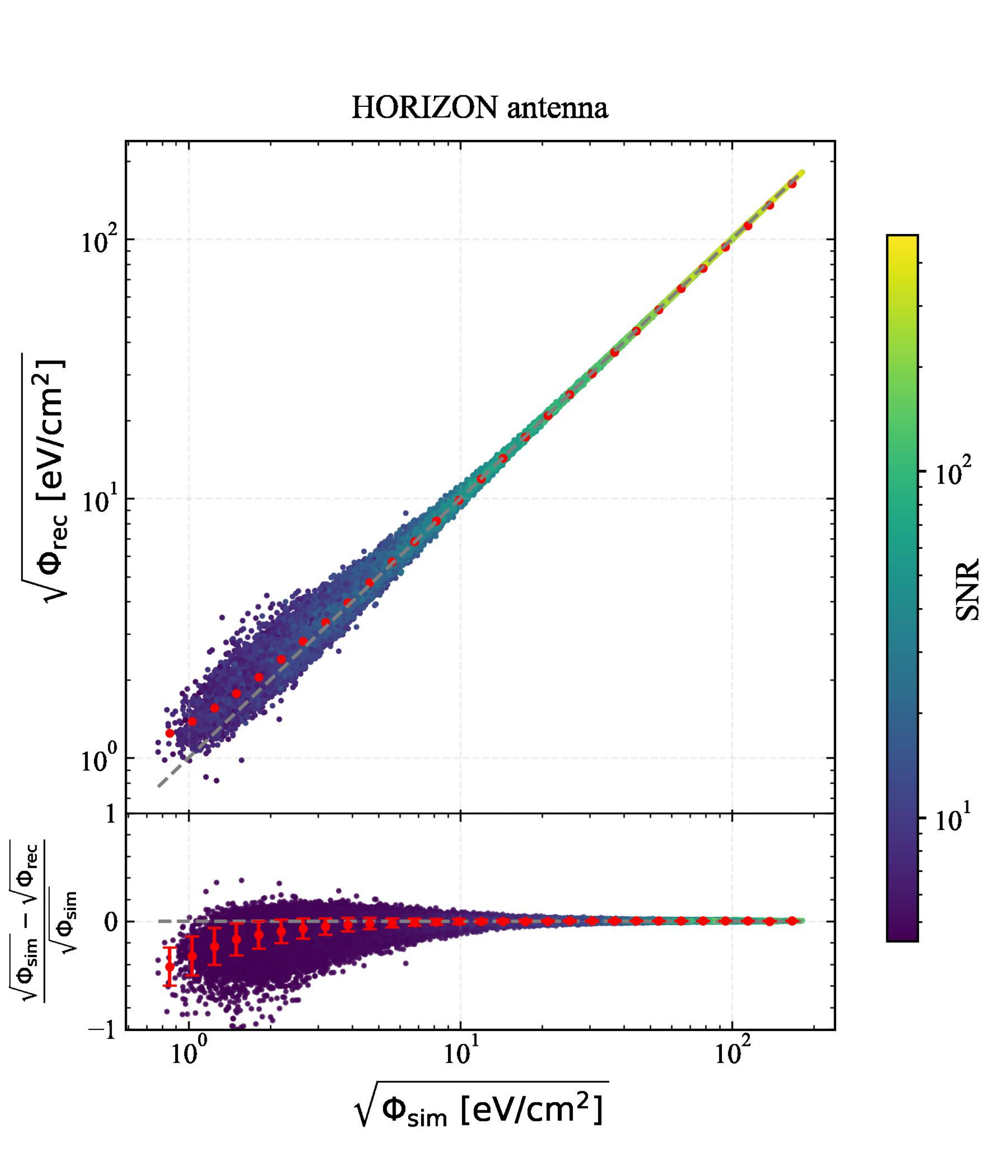}

\caption{Square root of the reconstructed energy fluence of the radio signals from simulated iron and proton air showers. The comparison contrasts the reconstructed energy fluence obtained via the analytical $\chi^2$ minimization method with the simulated energy fluence. The grey dashed line represents the true value, while the red points denote the mean values in each bin. The error bars in the bottom panel show the standard deviation within each bin. The color of each point indicates the SNR of the associated trace. Left: Results for the dipole antenna. Right: Results for the HORIZON antenna.}
\label{fig:fluence_snr}
\end{figure}

 For the dipole antenna, systematic underestimation persists, with mean and median values remaining both at 3\%. The total energy fluence exhibits a 4\% standard deviation, with a 68\% confidence interval ($\psi_{68}$) spanning [–0.04, 0.04]. For the HORIZON antenna, smaller systematic biases are observed: the mean relative error is null (0), while the median relative error is –0.02 – consistent with the bias documented in the analysis of the PEA. A 4\% standard deviation is observed, with $\psi_{68}$ spanning [–0.04, 0.01].
Consistent with the analysis of the PEA, the standard deviation in energy fluence for the $\theta$-component substantially exceeds that of the $\varphi$-component across both antenna types, attributable to the comparatively weaker signal strength in the former.
Despite these disparities, for both types of antennas the lsq reconstruction achieves high precision, thereby validating its methodological robustness.

Figure \ref{fig:fluence_snr} compares the reconstructed and simulated square root of energy fluence $\sqrt{\Phi} $, which is proportional to  E$_\mathrm{em}$. For each event, the SNRs at the antennas vary depending on the distance and viewing angle relative to the shower maximum. Additionally, for each antenna, the sensitivity of the three arms also depend on the air shower direction, leading to SNR differences in each channel. We classify events according to SNR, and we find that our analytical $\chi^2$ minimization yields accurate energy fluence estimates at high SNRs ($>$10) for both types of antennas.  However,  at lower SNRs, larger deviations are observed, indicating greater uncertainties in the reconstruction of shower energy E$_\mathrm{shower}$, with a noticeable bias that depends on the antenna response, which should be noticed and be further calibrated when dealing with real experimental data.

\section{Conclusions and outlook}


In this study, we developed a new reconstruction approach based on the analytical $\chi^2$ minimization, tailored to improve the accuracy and robustness of electric field reconstruction — an initial and critical step that directly influences the precision of subsequent analyses. The reconstruction method proposed does not make any assumptions about signal characteristics, polarization features, the energy contribution from the different emission effects, or any other signal-related aspects.



We demonstrate that the conventional matrix inversion method often struggles to accurately reconstruct the electric field — especially when using all three polarizations — due to underestimated uncertainties in frequencies where the antenna gain is low or the signal-to-noise ratio is poor. These reconstruction uncertainties can propagate during the integration process used to estimate the radiation energy, ultimately affecting the precision of the reconstructed air shower energy. The analytical $\chi^2$ minimization method developed in this study effectively addresses this issue. Without relying on assumptions about signal characteristics, it enables robust electric field reconstruction, achieving a typical standard deviation of about 3\% for the PEA and around 5\% for the energy fluence. Despite the different antenna response characteristics of the dipole and HORIZON antennas, both contain three polarizations and show consistently strong statistical performance with this method. These results confirm its suitability for precision radio-based energy reconstruction in air shower studies.

Furthermore, we confirm that when the analytical $\chi^2$ minimization method is adopted, incorporating a vertical polarization into the antenna design significantly improves the reconstruction accuracy of the peak envelope  amplitude for both antenna configurations. This result demonstrates the superiority of using the analytical $\chi^2$ minimization method for accurate electric field reconstruction with three-polarization antennas. 

To assess the directional dependence of this method, we produced a 2D map of zenith and azimuth angles, colored by $W_{\psi_{68}}$. The results shows that, across the full angular range, both antenna types achieve a $W_{\psi_{68}}$ better than 4\%, demonstrating the robustness of the method in different directions.

The arrival direction of the signal serves as an input parameter in the first step of this reconstruction. 
Therefore we explored the influence of directional accuracy on the analysis. A Gaussian deviation of 1$^\circ$ was set to estimate this influence. For the dipole antenna, the relative error of the PEA of the reconstructed electric field exhibits minimal variation, as illustrated in Figure \ref{fig:dir_smearing}. However, for the HORIZON antenna, there is a larger relative error for air showers with a zenith angle greater than 85$^\circ$.  This reflects the impact of prominent changes in the antenna response on the $1^\circ$ scale for specific arrival directions on the reconstruction, providing a reference for our future optimization of the antenna design. 
In practice, this has no significant impact on the overall resolution, since the  number of such cases is statistically small.


An iterative strategy, in which the reconstructed air shower direction is used again as the input for the electric field reconstruction, could be employed to repeat the direction reconstruction and further improve this method.

This work is based on a ZHAireS air shower simulation library, which includes iron and proton air showers, and the response of an antenna to generated electric fields. Before the reconstruction, the galactic noise simulated with LFmap was added to the signal, in an attempt to make the simulations as close to an ideal detection scenario as possible.  We have thus shown that, as long as the local noise is well understood and under control, this method is applicable in radio detection. Further tests to the method using the noise data measured by a real antenna array at an experimental radio site should be made to asses its performance in more challenging conditions. 

We tested the frequency band of \unit[30$-$200]{MHz}, which is covered by several experiments. For experiments beyond this range, further research is still needed. Within the frequency band we are studying, we did not subdivide it into sub-bands for detailed investigation. This aspect could be further explored in the future as, depending on the antenna response pattern, there could be directions where the accuracy of the method could be improved by restricting the signal to a specific sub-band.


\acknowledgments
We would like to thank Kaikai Duan, Xiaoyuan Huang, Jiale Wang, Jean-Marc Colley and all the other GRAND members for the related discussions.  This work is supported by the National Natural Science Foundation of China (Nos. 12273114), the Project for Young Scientists in Basic Research of Chinese Academy of Sciences (No. YSBR-061), and the Program for Innovative Talents and Entrepreneur in Jiangsu, and High-end Foreign Expert Introduction Program in China (No. G2023061006L). C.Z thanks Zhuo Li and Ruoyu Liu for their support and discussions during this work. 

\clearpage
\appendix
\section{Antenna response} 
\label{Responses}
\begin{figure}[ht]
\centering
\includegraphics[width=1.\textwidth]{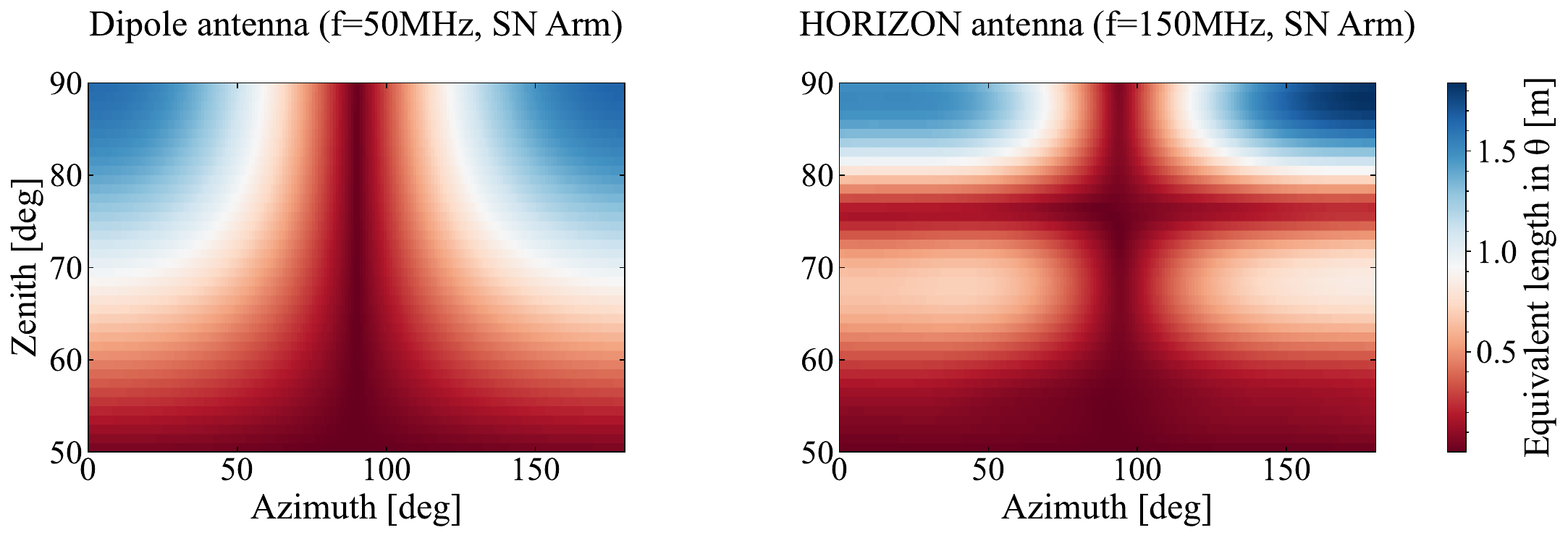}
    \caption{ Antenna response patterns of \(\theta\)-component at 50 MHz for the dipole antenna (left) and 150 MHz for HORIZON (right) antenna.  The radio signal is weaker in this component, and most errors arise from this component. }
    \label{fig:response_60MHz}
\hspace{1.6cm}
\includegraphics[width=1.\textwidth]{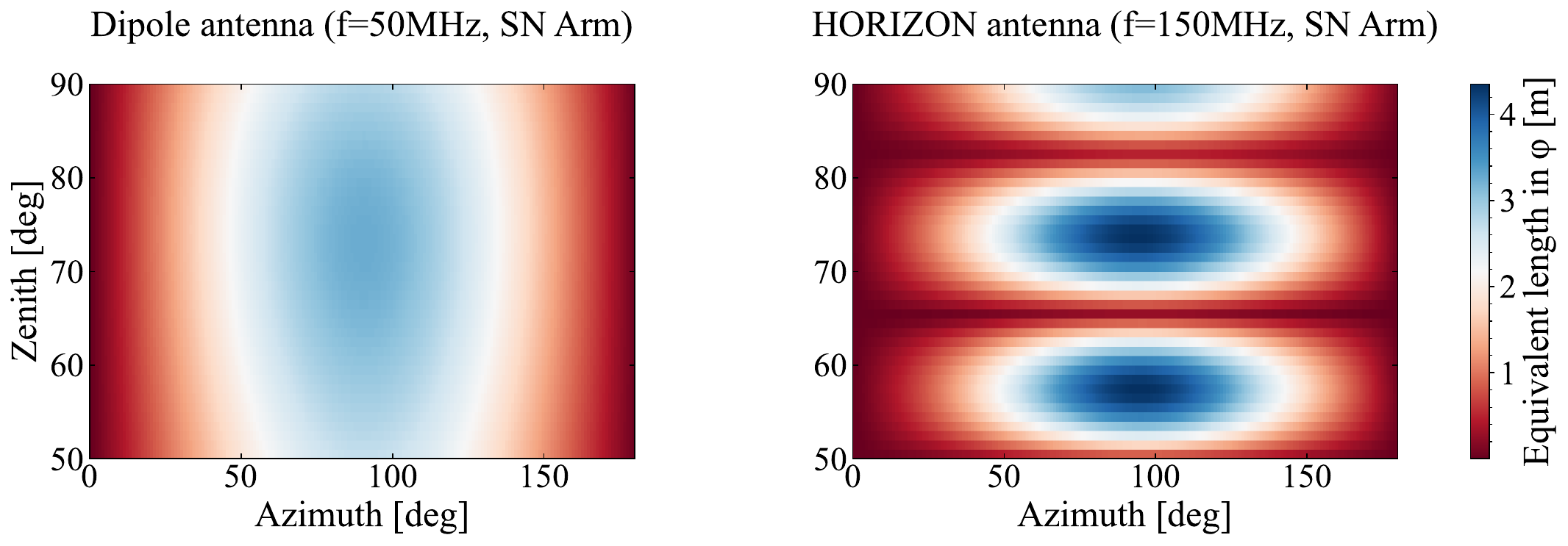}
    \caption{Antenna response patterns of the \(\varphi\)-component at \unit[50]{MHz} for the dipole antenna (left) and \unit[150]{MHz} for HORIZON (right) antenna.  The radio signal predominantly appears in this component, especially for inclined air showers.}
    \label{fig:response_60MHz_p_com}
\end{figure}

\clearpage
\section{Distribution of the relative error of the peak envelope amplitude}
\label{dir_fix}
\begin{figure}[htbp]
\centering
\includegraphics[width=0.95\textwidth]{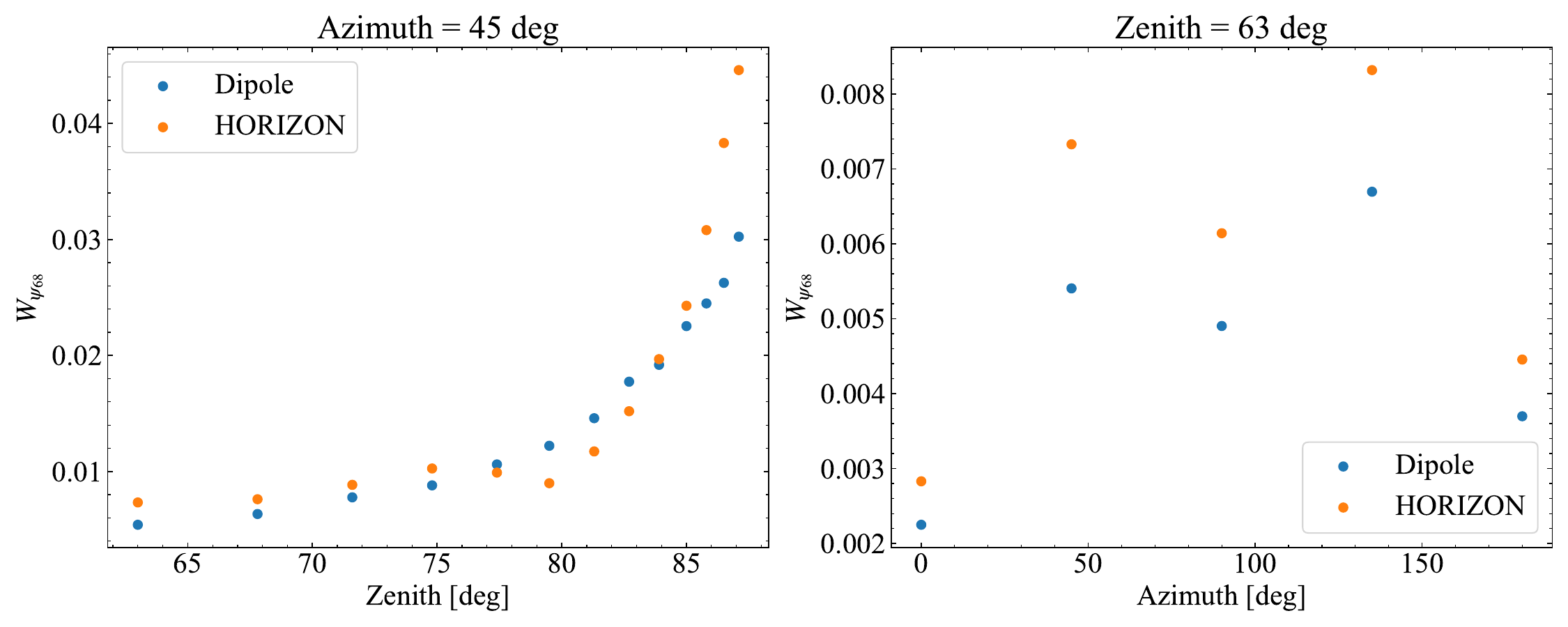}
    \caption{The total $W_{\psi_{68}}$  of the  distribution of relative error of the PEA of the reconstructed electric field as function of zenith (at azimuth = 45$^\circ$ (left)) and azimuth (at zenith = 63$^\circ$ (right)) angle for the HORIZON (orange)  and dipole (blue) antennas.}
    \label{fig:dir_dep_fix}
\end{figure}

\section{Peak envelope amplitude obtained with the matrix inversion method}
\label{1d_dir_dep}
\begin{figure}
    \centering
    \includegraphics[width=1\linewidth]{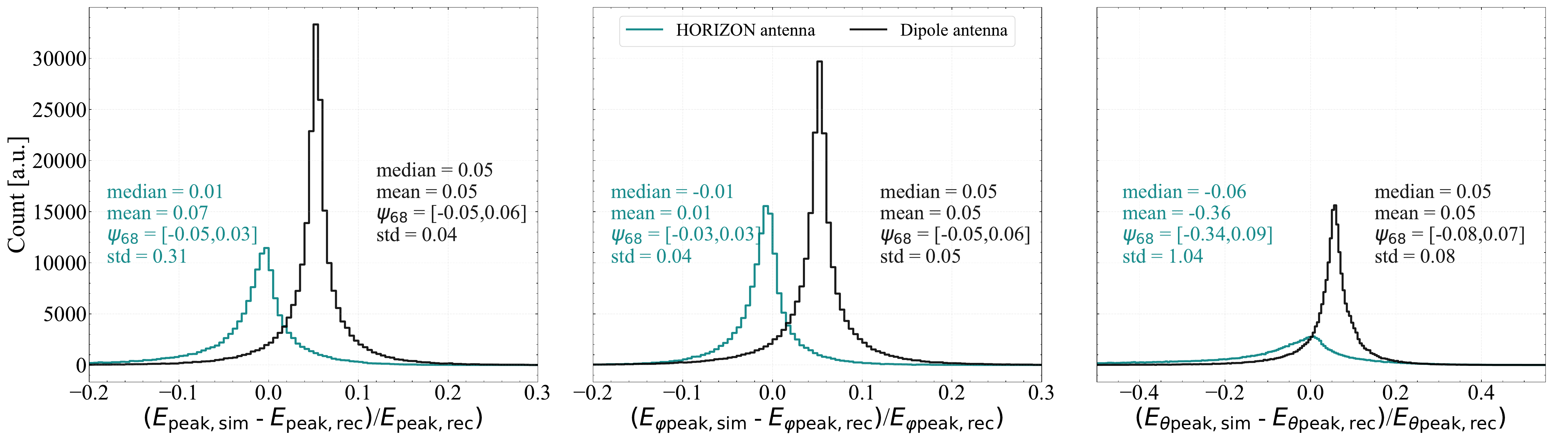}
    \caption{The distribution of relative error in the PEA reconstructed using the matrix inversion method for the total value(left), and the $\varphi$  (middle)  and $\theta$ (right) components for the HORIZON antenna (green) and the dipole (black) antenna.}
    \label{fig:e_rec_inv}
\end{figure}

Figure \ref{fig:e_rec_inv} presents the distribution of relative error in the PEA obtained through the Hilbert transform reconstructed using the matrix inversion method with three polarizations. While the results for the dipole antenna remain similar to those obtained with the analytical $\chi^2$ minimization method, the performance of this method worsens significantly for the HORIZON antenna. Specifically, the distribution of results becomes more spread, especially in the $\theta$-component, leading to greater uncertainty. The matrix inversion method struggles to accurately handle complex antenna responses, resulting in less reliable reconstructions.

\clearpage

\end{document}